\documentclass{emulateapj}
\usepackage{amsmath}
\usepackage{graphicx}
\usepackage{epstopdf}
\usepackage{hyperref}
\usepackage{xcolor}

\long\def\symbolfootnote[#1]#2{\begingroup%
\def\thefootnote{\fnsymbol{footnote}}\footnote[#1]{#2}\endgroup}

\newcommand{\beq}{\begin{equation}}
\newcommand{\eeq}{\end{equation}}
\newcommand{\bea}{\begin{eqnarray}}
\newcommand{\eea}{\end{eqnarray}}

\newcommand{\ddr}{{\partial \over \partial R} }

\newcommand{\ddt}{{\partial \over \partial t} }

\newcommand{\dMdr}{{dM_d\over d R}}
\newcommand{\dMdotwdr}{{d \dot M_w\over d R}}
\newcommand{\Sigmadotw}{\dot{\Sigma}_w}
\newcommand{\Mdot}{\dot{M}}

\slugcomment{Submitted to ApJ}

\shorttitle{TDE Disk Evolution}
\shortauthors{Shen \& Matzner}

\begin{document}

\title{Evolution of Accretion Disks in Tidal Disruption Events}

\author{Rong-Feng Shen\altaffilmark{1} and Christopher D. Matzner}
\email{rf.shen@mail.huji.ac.il}
\email{matzner@astro.utoronto.ca}
\affil{Department of Astronomy \& Astrophysics, University of Toronto, M5S 3H4, Canada}
\altaffiltext{1}{Current address: Racah Institute of Physics, Hebrew University of Jerusalem, Israel}

\begin{abstract}
During a stellar tidal disruption event (TDE), an accretion disk forms as stellar debris returns to the disruption site and circularizes. Rather than being confined within the circularizing radius, the disk can spread to larger radii to conserve angular momentum. A spreading disk is a source of matter for re-accretion at rates which can exceed the later stellar fallback rate, although a disk wind can suppress its contribution to the central black hole accretion rate. A spreading disk is detectible through a break in the central accretion rate history, or, at longer wavelengths, by its own emission. We model the evolution of TDE disk size and accretion rate, by accounting for the time-dependent fallback rate, for the influence of wind losses in the early, advective stage, and for the possibility of thermal instability for accretion rates intermediate between the advection-dominated and gas-pressure dominated states. The model provides a dynamic basis for modeling TDE light curves. All or part of a young TDE disk will precess as a solid body due to Lense-Thirring effect, and precession may manifest itself as quasi-periodic modulation of light curve. The precession period increases with time. Applying our results to the jetted TDE candidate Swift J1644+57, whose X-ray light curve shows numerous quasi-periodic dips, we argue that the data best fit a scenario in which a main-sequence star was fully disrupted by an intermediate mass black hole on an orbit significantly inclined from the black hole equator, with the apparent jet shutoff at t= 500 d corresponding to a disk transition from the advective state to the gas-pressure dominated state.
\end{abstract}

%\date{}


\maketitle

%%%%%%%%%%%%%%%%%%%%%%%%%%%%%%%%%%%%%%%%%%%%%%%%%%%%%%%%%%%%%%%%%%%%
\section{Introduction}

The tidal disruption of stars, first investigated as a primary means to grow supermassive black holes (e.g., Hills 1975), has more recently gained interest as a way in which the  $10^6 - 10^8 M_{\odot}$ black holes (BHs) in non-active galaxies may signify their existence (e.g., Rees 1988, 1990; Phinney 1989; Evans \& Kochanek 1989).

With the rapid advancement in the time-domain astronomy, stellar tidal disruption events (TDEs) have received increasing attention. So far a dozen or so TDE candidates have been observed. They were detected in X-ray bands early on, e.g., by XMM-Newton (Esquej et al. 2008), and more recently in UV / optical wavebands as well, e.g., by GALEX, Palomar Transient Factory and Sloan Digital Sky Survey (Gezari et al. 2008, 2009, 2012; 	van Velzen et al. 2011; Cenko et al. 2012a). The recent Swift X-ray transient Sw J1644+57 is the clearest TDE candidate so far.  The duration of this X-ray transient, and its location inside the host galaxy, are consistent with predictions for a tidal disruption flare (Barres de Almeida \& De Angelis 2011; Levan et al. 2011; Bloom et al. 2011; Burrows et al. 2011; Zauderer et al. 2011; Krolik \& Piran 2011; although an alternative interpretation does exist, e.g., Quataert \& Kasen 2012). Its peculiar emission properties imply this event has relativistic jet (Giannios \& Metzger 2011; Metzger, Giannios \& Mimica 2012; Berger et al. 2012; Zauderer et al. 2013). A second possibly jetted TDE candidate, Swift J2058+0516, was also discovered (Cenko et al. 2012b).

Given these developments, it is appropriate to review and improve the theoretical models connecting TDEs and their observables. Past modeling of TDE flares has often assumed the accretion rate onto the BH is identical to the rate at which
bound debris falls back to its periasteron and circularizes (e.g., Lodato, King \& Pringle 2009; Strubbe \& Quataert 2009; Lodato \& Rossi 2011; Krolik \& Piran 2012; Haas et al. 2012). We shall find, however, that the orbiting relic of early, rapid accretion -- a structure which can expand well beyond the disruption radius under certain circumstances -- is a potentially greater source of matter for later accretion on to the black hole.   Though its contribution can be suppressed by a wind, when it exists this `spreading disk' is guaranteed to become the predominant source of central accretion at sufficiently late times.

A change in the decay rate of central accretion is therefore one observable consequence of the spreading disk's existence, but not the only one.
Because it is a store of angular momentum, its presence affects the rate at which the disk undergoes Lense-Thirring precession.   Precession is a plausible explanation of the evolving quasi-periodic modulation of the Sw J1644+57 light curve (as previously considered by Stone \& Loeb 2012 and Lei, Zhang \& Gao 2013), so we are motivated to re-examine this issue in the context of a spreading disk.  A disk at large radii will also be detectable due to its own emission, especially at longer wavelengths which sample larger radii, but we defer this calculation to a companion paper (Shen et al. 2014, in prep.).

Our goals are to develop a comprehensive theory for the disk evolution from its inception to late times; to explore the dependence of this model on the parameters we use to describe viscosity and wind emission; and to predict the time evolution of the Lense-Thirring precession rate.   We review the parameters of stellar disruption in \S \ref{S:TDEbasics} before addressing the physical states of TDE accretion disks (\S \ref{sec:disk-eqns}), their evolution (\S \ref{sec:no-fallback} and \S \ref{sec:with-fallback}), and their precession (\S \ref{S:Precession}).    We rely for our analysis on two appendices: a new, self-similar treatment of a spreading disk which emits a wind (Appendix \ref{Sec:app-ss}), and a calculation of wind-free disk evolution with time variable fallback mass supply (Appendix \ref{sec:app-green}). 

Critically, we shall assume that an advective disk emits an unbound wind; see Loeb \& Ulmer (1997) and Coughlin \& Begelman (2013) for the alternative scenario in which the hole is enshrouded by weakly bound matter.

Our analysis is not, of course, without precedent, considering that spreading disks are a basic consequence of angular momentum conservation (Pringle 1981). Cannizzo, Lee \& Goodman (1990) have previously studied the viscous evolution of the TDE disk, but considered only the radiative, gas pressure dominated phase which sets in decades after the disruption.  Montesinos Armijo \& de Freitas Pacheco (2011) have also considered viscous evolution, but only the very earliest times of order the initial viscous time scale.

%%%%%%%%%%%%%%%%%%%%%%%%%%%%%%%%%%%%%%%%%%%%%%%%%%%%%%%%%%%%%%%%%%%%
\section{Tidal disruption of a star} \label{S:TDEbasics}

When an unlucky star plunges too close to a supermassive black hole, such that its pericenter distance $R_p$ is inside its tidal disruption radius
\beq
R_t = R_* (M/M_*)^{1/3}= 23~ M_6^{-2/3} m_*^{-1/3} r_* ~R_
{S},
\eeq
but outside a minimum radius which is slightly beyond $R_{S}$ (Darwin 1959), the BH's tidal force exceeds the star's self gravity and tears it apart, but does not immediately consume it.  Here $M= 10^6 M_6 M_{\odot}$ and $R_{S}= 2 G M/c^2$ are the BH's mass and Schwarzschild radius, and $M_*= m_* M_{\odot}$ and $R_*= r_* R_{\odot}$ are mass and radius of the star, respectively.    The depth of star's plunge is described by $\beta= R_t/R_p$, where $R_p$ is the pericenter radius; tidal disruptions occur when $1 \lesssim \beta \lesssim R_t/R_S$.

If the star is fully disrupted about half its mass becomes bound to the black hole, and the most tightly bound matter returns after a lag $t_{\rm ret}$ from the pericenter passage.  As we are interested in the dynamics of gas after it returns, we define $t=0$ at this point
\footnote{We note, however, that there can exist a prompt emission signal at $t \simeq -t_{\rm ret}$ in a deep plunging event, e.g., those associated with the shock breakout following the tidal compression of the star (Kobayashi et al. 2004; Guillochon et al. 2009), or when relativistic effects induce early accretion, as is seen in deep encounters of white dwarfs with intermediate massive black holes (Haas et al. 2012).}, so the star is disrupted at $t=-t_{\rm ret}$.   Two-fifths of the remaining bound matter, or about $M_*/5$, then arrives over a characteristic fallback time $t_f$, which is comparable to $t_{\rm ret}$.   However,  the least-bound portions trickle back much later:  for a uniform distribution of mass per unit specific energy across zero energy, the rate of fallback declines as $t^{-5/3}$ at very late times. (The same power law arises, for the same reason, in neutron star accretion of low-pressure ejecta during supernovae: Michel 1988, Chevalier 1989.)

If $t_*$ is some characteristic return time, the rate of fallback can therefore be described by $\dot M_{\rm fb} =M_*/(2t_*) {\cal G}(t/t_*)$ where the dimensionless fallback rate ${\cal G}(x)=0$  for $x<0$ and ${\cal G}(x)\propto x^{-5/3}$ for $x\gg1$; from our definitions, $\int_0^{t_f/t_*} {\cal G}(x) dx=2/5$ and $\int_0^{\infty} {\cal G}(x) =1$.  The precise functional form of ${\cal G}(x)$ and the ratios $t_{\rm ret}/t_*$ and $t_f/t_*$, depend on the dimensionless parameters of the disruption -- primarily the penetration factor $\beta$, the distribution of density within the initial star,  and $M_*/M$, but also, for very deep plunges, the relativity factor $R_p/R_S$ and the spin parameters.  All of these functions can be determined from numerical experiments (e.g., Lodato et al.\ 2009, Ramirez-Ruiz \& Rosswog 2009).  Given its constraints, the simple approximation ${\cal G}(x>0)=(2/5)\min[1,(t_*x/t_f)^{-5/3}]$ is sufficiently accurate for our purposes.  In dimensional terms this corresponds to
\beq		\label{eq:dotmfb}
\dot{M}_{\rm fb}(t) \simeq \dot M_f \times \begin{cases}
0, & ~~ t<0, \\
1, & ~~ 0<t < t_{\rm f},\\
(t_f/t)^{5/3}, & ~~ t > t_{\rm f}	\end{cases}
\eeq
where $\dot M_f = M_*/(5 t_f)$.  In this approximation, the dynamics of the disruption and the ensuing fallback are encapsulated in the ratio $t_f/t_*$.

\begin{figure}
\centerline{
\includegraphics[width=8.5cm, angle=0]{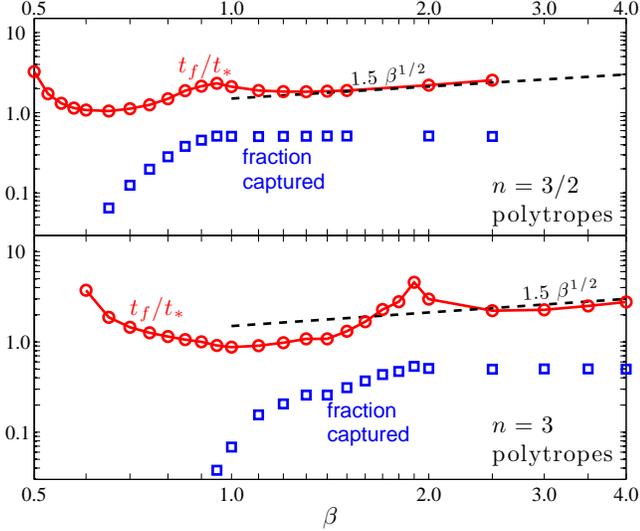}
}
\caption{Relation between the penetration factor $\beta$ and the characteristic fallback duration $t_f$ (the period over which 2/5 of bound matter returns) for $n=3$ (top) and $n=3/2$ (bottom) polytrope stars, in non-relativistic simulations with $M/M_*=10^6$ by Guillochon \& Ramirez-Ruiz (2013). In both cases, a power-law relation $t_f = 1.5\beta^{1/2} t_*$ is reasonably accurate, as is the statement that $2\lesssim t_f/t_* \lesssim 3$ for full disruptions with $\beta\lesssim4$.} \label{fig:Tf_Beta_Fit_RR13}
\end{figure}

For $t_*$ we adopt the period of a free orbit which is comoving with the star's center of gravity, but displaced inward by $R_*$ as it crosses the tidal radius:  
\beq \label{eq:tstar} 
\begin{split}
t_* &= \pi R_t^3/(2 G M R_*^3)^{1/2} \\ 
&= 40.5 ~ M_6^{1/2} r_*^{3/2} m_*^{-1} \, {\rm days}.
\end{split} 
\eeq
The relationship between $t_f$ and $t_*$ depends on dynamics -- that is, on the structure of the star and the parameters of the encounter.   Some analytical treatments assume the specific binding energy corresponds to an undistorted star at pericenter, which yields $t_f/t_* \propto \beta^{-3}$ (e.g., Evans \& Kochanek 1989; Ulmer 1999; Strubbe \& Quataert 2009; Lodato \& Rossi \ 2011). However, Stone, Sari \& Loeb\ (2012) argue, and recent simulations by Guillochon \& Ramirez-Ruiz (2013) verify, that $t_f/t_*$ is in fact insensitive to $\beta$ because the star is already disrupted somewhat inside $R_t$.
Relativistic effects such as black hole spin become important only in the deepest disruptions.\footnote{
During the encounter, tidal spin-up of the star (e.g., Li, Narayan \& Menou 2002) has a negligible effect on the spread of specific energy, its relative effect being $\sim (M_*/M)^{1/3} \ll 1$ (e.g., Evans \& Kochanek 1989).}
Accordingly, we adopt $t_f/t_*$ as a parameter.  This ratio is between 1 and 3 for a wide range of $\beta$, for polytropic stars of index $n=3/2$ or $n=3$, in the nonrelativistic simulations of Guillochon \& Ramirez-Ruiz (2013).  Figure \ref{fig:Tf_Beta_Fit_RR13} shows that $t_f/t_* = 1.5 \beta^{1/2}$ is a decent fit, but the range $2\lesssim t_f/t_*\lesssim 3$ describes full disruptions with $\beta\lesssim 4$. 

%%%%%%%%%%%%%%%%%%%%%%%%%%%%%%%%%%%%%%%%%%%%%%%
%%%%%%%%%%%%%%%%%%%%%%%%%%%%%%%%%%%%%%%%%%%%%%%%%%%%%%%%%%%%%%%%%%%%%%%%%%%%%%%%%

\begin{figure}
\centerline{
\includegraphics[width=9cm, angle=0]{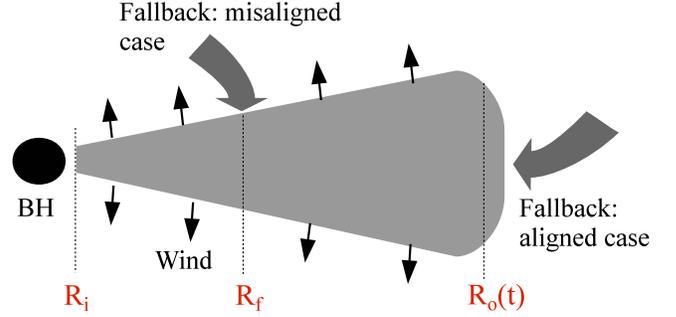}
}
\caption{Sketch of a TDE accretion disk with fallback and wind mass loss in the early, advection-dominated phase of a disruption event. In the case that the black hole spin is aligned with the normal of the stellar orbit plane, the infall material joins the disk at the outer disk radius $R_o$. In the misaligned case, the precession removes the disk from the infall plane, so that new matter arrives at $R_f$ rather than $R_o$. Note that a wind may also be launched from the site where the infall material joins the disk due to shock heating (Strubbe \& Quataert 2009) which reduces the rate at which fallback mass joins the disk.} \label{fig:disk}
\end{figure}

Within a few orbits, the returning bound debris material collides with itself, eventually settling at its circularization radius
\beq
R_f= 2 R_p= 47 ~\beta^{-1} M_6^{-2/3} m_*^{-1/3} r_* ~R_S
\eeq
before accreting onto the hole. Figure \ref{fig:disk} illustrates the TDE accretion disk with fallback.

Electron scattering dominates the opacity $\kappa$ and most of the disrupted stars will be of roughly Solar metallicity, so we take  the mean molecular weight to be $\mu= 0.6$ and adopt $\kappa = 0.34 $\,cm$^{2}$\,g$^{-1}$ throughout. Normalized to a critical accretion rate $\dot{M}_{\rm crit}= L_{\rm Edd}/c^2$ where $L_{\rm Edd}$ is the Eddington luminosity, the peak fallback rate is 
\beq
\dot{m}_{\rm f} \equiv \frac{\dot{M}_{\rm f}}{\dot{M}_{\rm crit}}= 690
~ (t_f/t_*)^{-1} M_6^{-3/2} r_*^{-3/2} m_*^2.
\eeq
The early, highly super-Eddington fallback rate implies that the disk will be radiatively inefficient for some time after the event (Rees 1988), and this has important implications for our analysis below.

Two effects are neglected in our expressions for $\dot{M}_f$ and $t_f$. One is the possibility of partial disruptions (e.g., when $\beta \lesssim 1$ or if the star contains a dense core), for which the mass fraction lost by the star during the grazing diminishes and the late fallback drops somewhat more steeply than $t^{-5/3}$ (Guillochon \& Ramirez-Ruiz 2013).  These authors' simulations show that stars are fully disrupted when the impact parameter is above some threshold: $\beta> 0.9$ for $n=3/2$ polytropes, and $\beta>1.8$ for $n=3$ polytropes (see Fig.~\ref{fig:Tf_Beta_Fit_RR13}).  Another is the effect of relativity in very deep plunges, for which $R_p$ approaches the innermost stable circular orbit of the black hole. Using relativistic orbits whose energies are calculated assuming an undisturbed star at pericenter, Kesden (2012) argues that relativistic effects at most halve $t_f$ and double $\dot{m}_f$ when compared to the analogous Newtonian orbit.  If the energy distribution is set closer to $R_t$, as it appears to be, then the effect will be less than a factor of two. Therefore the above expressions of $t_f$ and $\dot{m}_f$ can be considered valid for non-relativistic full disruptions ($\beta \gtrsim 1$), which are the focus of this paper, and valid within roughly a factor of 2 for relativistic disruptions.

%%%%%%%%%%%%%%%%%%%%%%%%%%%%%%%%%%%%%%%%%%%%%%%%%%%%%%%%%%%%%%%%%%%%%%%%%%%%%%%%

%%%%%%%%%%%%%%%%%%%%%%%%%%%%%%%%%%%%%%%%%%%%%%%%%%%%%%%%%%%%%%%%%%%%%%%%%%%%%%%%%

%%%%%%%%%%%%%%%%%%%%%%%%%%%%%%%%%%%%%%%%%%%%%%%%%%%%%%%%%%%%%%%%%%%%%%%%%%%%%%%%%%%%%%%%%%%%%%%%%%%%%%%%%
\section{Disk Physics and Viscous Evolution}	\label{sec:disk-eqns}

Our goal is to address the viscous evolution to the long-term evolution of TDE flares.  Before we make any detailed models, we pause now to show that this ingredient is potentially very important.   Consider a disk which evolves due to its internal kinematic viscosity $\nu \propto R^n$ (i.e., a function of $R$ only), such that the local viscous time is $t_\nu = (2/3)R^2/\nu \propto R^{2-n}$, and neglect (only for the moment) the influences of continuous debris fallback and outflow from the disk's surface.

Because angular momentum is conserved, and because the specific orbital angular momentum $j=(GMR)^{1/2}$ increases with radius, a disk whose matter drains onto a compact central object must also expand in radius.  In particular, if a thin ring of matter is added to the disk at radius $R_f$, then it will spread radially over a time $t_{\nu0} = t_\nu(R_f)$ and begin to drain onto the central object (Pringle 1981).  After a couple of these initial viscous times ($t>t_{\nu0}$), the disk settles into a self-similar, spreading state with outer radius $R_o(t)$ that expands to keep the viscous time $t_\nu(R_o)$ comparable to its age, so $R_o\propto t^{1/(2-n)}$.  Angular momentum conservation then requires that the disk mass decline as $M_d \propto R_o^{-1/2}\propto t^{1/(4-2n)}$, and the central accretion rate decline as $\dot M_{\rm acc} \propto {M_d/ t} \propto t^{-\eta}$ with $\eta = (5-2n)/(4-2n)$.
So long as $\nu$ depends only on $R$, any matter added later undergoes precisely the same evolution, offset in time, which adds linearly to the disk surface density $\Sigma(R,t)$ and the central accretion rate $\dot M_{\rm acc}(t)$.   In a TDE, $t_{\nu0}\lesssim t_f$, so the early viscous time is not a significant delay.  We explore this scenario further in Appendix~\ref{sec:app-green} by means of Green's function.

In the late phases of a TDE new stellar matter continues to fall back, at the diminishing rate $\dot M_{\rm fb}\propto t^{-5/3}$.   Critically, however, it is possible for the disk accretion rate to decline more slowly.  In the example just given, this occurs when $n<5/4$, and typical values of $n$ are indeed below $5/4$: see \S\ref{SS:viscosity}.

Even for a more general case in which $\nu$ is a power law function of not only $R$ but also $\Sigma$, i.e., $\nu \propto \Sigma^q R^n$, there
exists a self-similar spreading solution for which $t_\nu(R_o)/t$ remains constant and
\beq		\label{eq:pringle}
\eta= \frac{5q+5-2n}{5q+4-2n}~~~~~~(\rm{no~wind})
\eeq
so long as $q\geq \max(0, n/2-1)$ (Pringle 1991).
 As we will see below,  for values of $q$ and $n$ relevant to TDEs, the central disk accretion rate always declines slower than $t^{-5/3}$ unless its evolution is affected by a disk wind.

In other words, the spreading remnant of early fallback has the potential to overwhelm the returning stream of stellar matter as a source of accretion onto the central object at late times in TDEs.  Even in cases or phases where this does not occur, the outer disk (when present) can signal its existence through its own emission at long wavelengths, by affecting the Lense-Thirring precession rate, or by emitting a wind. See Figure \ref{fig:disk} for illustration.

%%%%%%%%%%%%%%%%%%%%%%%%%%%%%%%%%%%%%%%%%%%%%%%%%%%%%%%%%%%%%%%%%%%%%%%%%%%%%%%%
\begin{figure}
\centerline{
\includegraphics[width=8.5cm, angle=0]{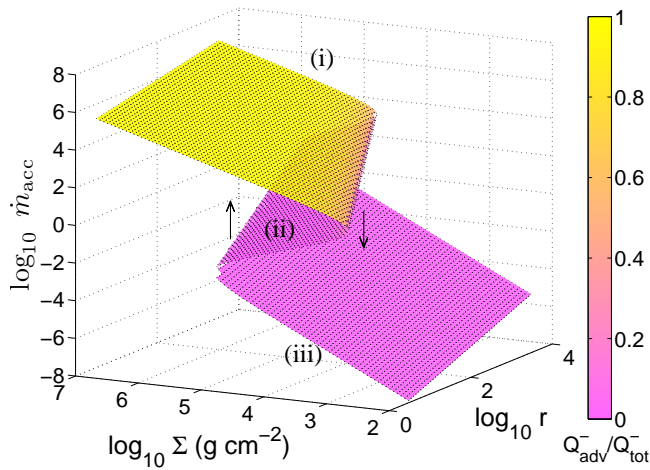}
}
\caption{The steady state disk solution numerically calculated from Equation (\ref{eq:energy}) for $M= 10^6 M_{\odot}$ and $\alpha=0.01$. The color coding is for the fraction of the total disk cooling that is carried by advection. Three regimes of the solution are visible: (i) high $\dot{m}_{\rm acc}$, $Q_{\rm adv}^-$- and $P_{\rm rad}$-dominated; (ii) intermediate $\dot{m}_{\rm acc}$, $Q_{\rm rad}^-$- and $P_{\rm rad}$-dominated, which is thermally and viscously unstable; (iii) low $\dot{m}_{\rm acc}$, $Q_{\rm rad}^-$- and $P_{\rm gas}$-dominated. The arrows indicate the directions of regime transitions between (i) and (iii) relevant to TDEs (\S\ref{SS:StateTransitions}).}\label{fig:scandisk}
\end{figure}
%%%%%%%%%%%%%%%%%%%%%%%%%%%%%%%%%%%%%%%%%%%%%%%%%%%%%%%%%%%%%%%%%%%%%%%%%%%%%%%%%

\subsection{Disk structure} \label{SS:DiskStructure}

In order to understand the disk's evolution, we need to determine its structure, especially the radial dependence of the column density $\Sigma$ and the viscosity $\nu$, i.e., the values of $q$ and $n$. The rest of the section fills in the physical details required to arrive at a quantitative model of the spreading disk. We will consider both advective and radiative coolings. Our procedure is similar to those of Cannizzo \& Gehrels (2009) and Strubbe \& Quataert (2009), but we focus on the physical state changes in the long-term evolution of disk, and we self-consistently consider an intermediate phase during which the disk crosses an unstable branch of its evolution track.

The disk half thickness is $H= c_s/\Omega_k$, where $c_s= (P/\rho)^{1/2}$ is the isothermal sound speed and $\Omega_k$ is the disk angular speed which we assume to be Keplerian. The pressure is the sum of the radiation pressure and the gas pressure: $P= P_{\rm rad}+ P_{\rm gas}$ $= a T^4 /3 + \rho k T/(\mu m_p)$. The disk surface density is defined as $\Sigma= \int_{-\infty}^{\infty} \rho dz= 2\rho H$.
Assuming steady state accretion without infall or accretion leads to the standard relations  $v_r= 3\nu/(2R)$ and $\dot{M}_{\rm acc}= 2\pi R \Sigma v_r= 3\pi \nu \Sigma$.  However, these expressions are modified when matter arrives or is emitted with non-Keplerian angular momentum; see Equation  (\ref{eq:vR-general}) of Appendix \ref{Sec:app-ss}.

We consider a steady-state disk, for which the heating is balanced with the cooling at each radius. The viscous heating rate per unit surface area of the disk is $Q^+= \nu \Sigma R^2 (\partial \Omega_k/\partial R)^2= 9 \nu \Sigma \Omega_k^2 /4$.  The advective cooling rate for the same region is $Q^-_{\rm adv}= \Sigma v_r T (\partial s/\partial R) \simeq \Sigma v_r P/(\rho R)$, where $s$ is the entropy per unit mass, and in writing the second step we neglected a numerical coefficient of order unity (e.g., Kato et al. 1998). The radiative cooling rate from the two faces of disk is $Q^-_{\rm rad}= 4 a c T^4/(3 \kappa \Sigma)$, where $\kappa$ is the opacity which in the TDE context is dominated by free electron scattering. Additionally, when the disk is in the high accretion rate regime where $Q^-_{\rm adv}$ dominates over $Q^-_{\rm rad}$, a fraction of disk mass is likely to be unbound and blown off in a disk wind, so that the local accretion rate decreasing inward as $\dot{M_{\rm acc}} \propto R^s$; wind carries away some energy. Therefore, the energy equation reads as $Q^+= Q^-_{\rm adv}+Q^-_{\rm rad} + Q^-_{\rm w}$. However, as long as $s$ is constant, $Q^-_{\rm w}$ is always a constant fraction of $Q^+$, one that vanishes when there is no wind  (see Equation \ref{eq:wind-energy}).
Thus, for our purposes here, $Q^-_{\rm w}$ can be dropped and the energy equation is written as
\beq		\label{eq:energy}
\frac{9}{4} \nu \Sigma \Omega_k^2 \simeq \frac{\dot{M}_{\rm acc}}{2\pi R^2} \frac{P}{\rho} + \frac{4 a c T^4}{3 \kappa \Sigma}.
\eeq

From Equation (\ref{eq:energy}) one can identify a few limiting accretion regimes, and then find radial dependeces of $\nu$ and $\Sigma$, in turn the disk temporal behavior for each regime. Before we delineate these regimes, we must address the form of $\nu$.

%%%%%%%%%%%%%%%%%%%%%%%%%%%%%%%%%%%%%%%%%%%%%%%%%%%%%%%%%%%%%%%%%%%%%%%%%%%%%%%%%%%%%
\subsection{Form of the viscosity law} \label{SS:viscosity}

Analytical models such as ours have traditionally relied either on the viscosity model of Shakura \& Sunyaev (1973) in which $\nu = 2\alpha P /(3 \Omega_K \rho)$, where $\alpha = 10^{-2} \alpha_{-2}$ is assumed to be reasonably constant, or on the revised model of Sakimoto \& Coroniti (1981) in which the gas pressure $P_{\rm gas}$ replaces the total pressure $P$; intermediate expressions are also possible.  The two prescriptions behave very differently when radiation pressure is significant, especially when radiative cooling is also important. Recent numerical simulations have shed important light on how these idealizations compare with the dynamics of the magneto-rotational instability in this regime.   In this section we shall first consider the behavior of the Shakura \& Sunyaev model, then contrast it with the Sakimoto \& Coroniti model, before addressing these numerical results.   We do not consider the original amplification of stellar magnetic fields to their saturated values, although we recognize that this merits closer scrutiny.

For convenience, from now on, we choose lowercase symbols to define the normalized mass rates $\dot{m}_{\rm acc} = \dot{M}_{\rm acc}/\dot{M}_{\rm crit}$, $\dot{m}_{\rm fb}= \dot{M}_{\rm fb}/\dot{M}_{\rm crit}$ and radius $r = R/R_S$, and use them where it is necessary (recall, however, that $r_*$ and $m_*$ are normalized to Solar values).  

Along a trend of decreasing $\dot{m}_{\rm acc}$, the cooling term at a given radius will first be dominated by advection and later by radiation; the total pressure is dominated by radiation early on, and later by gas pressure.  Figure \ref{fig:scandisk} plots the numerical solution to Equation (\ref{eq:energy}) in the $\dot{m}_{\rm acc}$-$\Sigma$-$r$ space, using the Shakura \& Sunyaev model with fixed $\alpha$.   It has three physical regimes: (i) high $\dot{m}_{\rm acc}$, advective cooling, radiation pressure; (ii) intermediate $\dot{m}_{\rm acc}$, radiative cooling, radiation pressure; (iii) low $\dot{m}_{\rm acc}$, radiative cooling, gas pressure.  Regime (i) corresponds to the ``slim disk'' model in the literature (e.g., Abramowicz et al. 1988), whereas regime (iii) is the standard Shakura \& Sunyaev disk.

In the advective regime (i), $Q^+ = Q^-_{\rm adv}$. From Equation (\ref{eq:energy}) one then easily finds
\beq
(H/R)_{\rm adv} \simeq 1,
\eeq
and $q=0$, $n=1/2$. The local viscous time scale is $t_{\nu}= (2/3) r^2/\nu= (\alpha \Omega_k)^{-1} (H/R)^{-2}$. What is useful is $t_{\nu0}$, the viscous time at $r_f$ in this regime, which we find to be
\beq    \label{eq:tnu0}
{t_{\nu,0}\over t_f }= 0.13~ \frac{(t_f/t_*)^{-1}}{\alpha_{-2} \beta^{3/2}} \left(m_*\over M_6\right)^{1/2}.
\eeq

In the radiative, radiation pressure dominated regime (ii), $Q^+ = Q^-_{\rm rad}$ and $P = P_{\rm rad}$. One finds $q= -2$ and $n= 3/2$.  The equilibrium state is characterized by $H/R \simeq \dot m_{\rm acc}$, but as the disk is unstable (see below), this merely serves to divide those disks that heat towards state (i) from those which cool towards state (iii).

In the radiative, gas pressure dominated regime (iii), $Q^+ = Q^-_{\rm rad}$ and $P = P_{\rm gas}$. Thus, one finds
\beq
(H/R)_{\rm gas}= 3.9\times10^{-3}~ (\alpha_{-2} M_6)^{-1/10} \dot{m}_{\rm acc}^{1/5} r^{1/20},
\eeq
and $q= 2/3$, $n= 1$. The local viscous time scale in this regime is
\beq    \label{eq:tvis-gas}
t_{\nu, \rm gas}= 9.1\times10^7~ \alpha_{-2}^{-4/5} M_6^{6/5} r^{7/5} \dot{m}_{\rm acc}^{-2/5} ~{\rm s}.
\eeq

The border between regimes (i) and (ii), where $Q^-_{\rm adv}=Q^-_{\rm rad}= Q^+/2$ and $P=P_{\rm rad}$, is
\beq		\label{eq:adv=rad}
(\dot{m}_{\rm acc})_{\rm i-ii} = \frac{2}{\sqrt{3}} r,
\eeq
and the one between regimes (ii) and (iii), where $Q^+= Q^-_{\rm rad}$ and $P_{\rm rad}= P_{\rm gas}= P/2$, is
\beq		\label{eq:rad=gas}
(\dot{m}_{\rm acc})_{\rm ii-iii} = 8.4\times10^{-4}~ (\alpha_{-2} M_6)^{-1/8} r^{21/16}.
\eeq
As accretion rate drops with time in a long trend, the transition of disk from one regime to the other can happen, during which the scalings of $\dot{m}_{\rm acc}(t)$ and $H/R$ change.

It is well known that the radiatively cooled, radiation-pressure dominated regime (ii) of a disk with the Shakura \& Sunyaev viscosity law is thermally unstable (Lightman \& Eardley 1974; Shakura \& Sunyaev 1976; see Kato et al. 1998 for a review). This can be seen from $Q^+ \propto T^8$ while $Q^-_{\rm rad} \propto T^4$: any increase of $T$ relative to steady state leads to excess heating, making the disk even hotter, whereas any slight decrease of $T$ triggers runaway cooling. When this viscosity prescription is used within one-dimensional numerical simulations of radiative, radiation-pressure dominated disks -- with fixed disk outer boundary and mass feeding rate  -- one observes globally limit-cycle behavior (Honma et al. 1991; Szuszkiewicz \& Miller 2001; Ohsuga 2005, 2007; Li, Xue \& Lu 2007; \S \ref{SS:StateTransitions}) in which the accretion rate and disk scale height jump between the high $\dot{m}_{\rm acc}$, advective regime and the low $\dot{m}_{\rm acc}$, gas-pressure dominated regime. The duration of one cycle roughly corresponds to the outer viscous time.

Thermal instability can be suppressed with a change to the viscosity law, such as Sakimoto \& Coroniti's prescription $\nu\propto P_{\rm gas}/(\rho\Omega^2)$ which has frequently been adopted in studies of black-hole accretion (e.g., Milosavljevi{\'c} \& Phinney 2005, Tanaka \& Menou 2010, Haas et al.\ 2012).   In this model regime (iii) and the boundary between (ii) and (iii) are unaffected, but regime (ii), which is now thermally stable, is characterized by $(n,q) = (1, 2/3)$.  The advective regime (i) is also dramatically altered: it also has $(n,q)=(1,2/3)$, rather than $(1/2,0)$.

Very recently, numerical simulations have reached the level of sophistication required to address the physical interplay between the magnetorotational instability (MRI) and radiation-matter interaction which characterizes regime (ii).  Hirose et al.\ (2009) used the Zeus code (Stone \& Norman 1992) modified by Turner \& Stone (2001) to include radiation transport in the flux-limited diffusion approximation.    Jiang et al.\ (2013) simulate the same physical problem with the Athena code (Stone et al.\ 2008) augmented with a variable Eddington tensor radiation transport (Davis et al.\ 2012, Jiang et al.\ 2012).  Whereas Hirose et al.\ find radiative, radiation-dominated disks to be thermally stable, Jiang et al.\ observe runaway heating or cooling in every example.   The reasons for this difference in behavior are not yet clear, but as we regard the Jiang et al.\ simulations as more sophisticated, we are led to conclude that, insofar as thermal stability is concerned, Shakura \& Sunyaev's model is favored over Sakimoto \& Coroniti's.

In truth, many of the numerical details are not well matched by either model.  Jiang et al.\ (2013) report that the heating and cooling rates scale as powers of $P$ which are non-integer and which depend on the initial conditions; moreover, a delay between runaway heating or cooling suggests something other than linear instability.

Despite these differences, the presence of thermal instability renders the same outcome: the disk must diverge from its unstable equilibrium and stabilize  either in state (i) due to the effects of advective cooling, or in state (iii) due to finite gas pressure.  We can therefore accept the predictions of the Shakura \& Sunyaev prescription, so long as it remains accurate in the advective state (i) as well as the radiative state (iii).   We believe it does, because the deeply advective, radiation pressure dominated limit resembles a completely nonradiative disk with $\gamma=4/3$, and this case is known to show saturated MRI (Hawley et al.\ 2001) which is adequately described by a characteristic $\alpha$.   For these reasons we adopt the Shakura \& Sunyaev viscosity law, while sounding a note of caution that the details of state transitions are not likely to be captured perfectly and that $\alpha$ may differ between states (i) and (iii).

%%%%%%%%%%%%%%%%%%%%%%%%%%%%%%%%%%%%%%%%%%%%%%%%%%%%%%%%%%%%%%%%%%%%%%%%%%%%%%%%
\subsection{Implications of Thermal Instability} \label{SS:StateTransitions}

We are concerned with the evolution of disks of declining accretion rate which may or may not receive matter at their outer edge.  As we are unaware of any global simulations lacking a source at large radii, we adopt a simple prescription based on the notion that thermal readjustments are more rapid than viscous ones.   (An important caveat is that in state (i), the thermal time is not in fact much shorter than the viscous time.)  Once the initially advective disk crosses the border between regimes (i) and (ii),  Equation (\ref{eq:adv=rad}), we assume that its temperature, scale height, and accretion rate immediately contract to the radiative, gas-pressure dominated state (iii).  In the $\dot{m}_{\rm acc}$-$\Sigma$-$r$ space of Figure \ref{fig:scandisk}, the disk falls vertically off the ledge and lands on the gas-pressure dominated regime.    If the disk was expanding self-similarly in the advective regime, so that $t_\nu(R_o)\simeq t$ in state (i), then the viscous time must suddenly become much greater than $t$ (Equation \ref{eq:tvis-gas}).

The subsequent evolution depends on the presence and rate of fallback supplying matter at the outer disk.  If there is none, it will stay in regime (iii).  Over the course of one viscous time its $\dot m_{\rm acc}$ will remain constant, but afterward it will follow the self-similar viscous behavior for $q$ and $n$ characteristic for this regime.  If instead there is continuous fallback at a rate characteristic of regime (ii), a limit cycle results.  Because of its long viscous time, the disk accumulates mass. It will move up in regime (iii) with increasing $\Sigma$, until it reaches the border of regimes (ii) and (iii), i.e., Equation (\ref{eq:rad=gas}). Then it will jump up directly to the advective regime (i). Because the mass depletion rate $(\dot{m}_{\rm acc})_{\rm adv}$ is so high, it stays in that regime only for a very short while before reaching the ledge again, then falling off to regime (iii), finishing one cycle.

Based on the above disk physics, we quantitatively describe the disk evolution in the next two sections.  Because most of the mass and angular momentum arrives within a few times $t_f$, and because disk precession can allow the late-arriving fallback to avoid colliding with the outer disk, we begin in  \S~\ref{sec:no-fallback} with the idealized case of a spreading disk where disk-driven winds are included, but fallback is entirely ignored.  To account for the influence of a wind from the nonradiative and accreting portions of the disk, we rely on the self-similar model for windy, spreading disks worked out in Appendix~\ref{Sec:app-ss}.  This provides a useful reference point for \S~\ref{sec:with-fallback}, where we consider the disk's evolution with fallback, and address two scenarios for the alignment of the disk and the black hole spin plane.  We then address the Lense-Thirring precession of the disk (\S~\ref{S:Precession}) and applying this to the event Sw J1644+57 (\S~\ref{S:SwJ1644}).

%%%%%%%%%%%%%%%%%%%%%%%%%%%%%%%%%%%%%%%%%%%%%%%%%%%%%%%%%%%%%%%%%%%%%%%%%%%%%%%%
\begin{figure}
\centerline{
\includegraphics[width=8.8cm, angle=0]{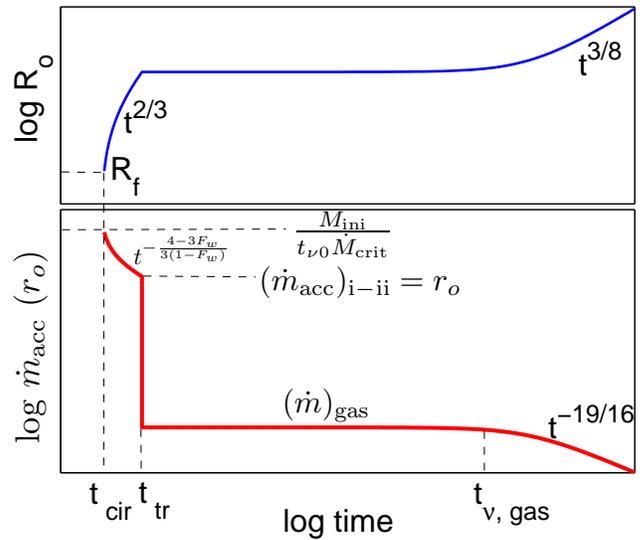}
}
\caption{Schematic evolution of the outer radius and accretion rate of a disk with an initial mass of $M_{\rm ini}$ but without later fallback. See \S \ref{sec:no-fallback} for explanation.}    \label{fig:no-fallback}
\end{figure}
%%%%%%%%%%%%%%%%%%%%%%%%%%%%%%%%%%%%%%%%%%%%%%%%%%%%%%%%%%%%%%%%%%%%%%%%%%%%%%%%%

%%%%%%%%%%%%%%%%%%%%%%%%%%%%%%%%%%%%%%%%%%%%%%%%%%%%%%%%%%%%%%%%%%%%%%%%%%%%%%%%%%%%%%%%%%%%%%%%
\section{Disk evolution without fallback}		\label{sec:no-fallback}

We begin with the question of what happens to a disk that was built up by stellar fallback matter over a few times $t_f$, but then receives no matter afterward.  For TDEs this is relevant as a limiting case, both because the timing of fallback adds most of the mass and angular momentum to the disk at early times, and because torque from the central object can swing the disk plane away from the fallback stream so that new infall arrives at $R_f$ rather than $R_o$.  However the expansion of an isolated, initially advective disk may be directly applicable to other physical problems such as the coalescence of compact binaries.

First we pause to consider how the debris stream circularizes to form a disk, which is, in fact, a complex processes. There are generally three effects (Evans \& Kochanek 1989; Kochanek 1994). First, at the pericenter, the compression shock due to an effective nozzle redistributes the angular momentum of returning material that passes through it.  Second, debris returns after a second or later pass and collides with more recently-arriving material. Third, relativistic precession causes the apsidal angle of the debris streams to precess, such that the outgoing gas is on an orbit that will collide with the ingoing gas. {Some of these effects are explored in simulations (e.g., Ramirez-Ruiz \& Rosswog 2009; Haas et al. 2012; Hayasaki, Stone \& Loeb 2012; Guillochon, Manukian \& Ramirez-Ruiz 2013) but a systematic investigation of the circularization process is still lacking.} One expects that generally the circularization would occur within a few fallback orbits, thus on a time scale of $t_{\rm cir}= n_{\rm cir} t_f$, with $n_{\rm cir}$ generally lying between 1 and 10. There should be no substantial accretion going on toward the black hole until the disk is set up. Right after $t_{\rm cir}$, the disk accumulated a mass of $M_{\rm ini}= \int_0^{t_{\rm cir}} \dot{M}_{\rm fb} dt$ at the fallback radius $R_f$, and this sets the initial viscous accretion rate $\approx M_{\rm ini}/t_{\nu0}$.
If $n_{\rm cir}$ is not large, there is little practical difference between evolving the circularized disk from $t=t_{\rm cir}$ with mass $M_{\rm ini}$, and evolving it from $t=0$ with zero mass.

Even without continued fallback, the disk's accretion rate falls from potentially super-Eddington values of order $\dot M_f$ towards zero, so we must consider both the advective phase and the transition to a gas pressure dominated, radiative phase, i.e., the transition from state (i) to state (iii) in the terminology of \S \ref{SS:viscosity}.  The evolution of a disk without fallback is shown in Figure \ref{fig:no-fallback}, which we now explain in detail.

To get the accretion rate evolution law we cannot use Equation (\ref{eq:pringle}) because an advective disk is likely to emit a wind. A  standard, if crude, treatment of the wind is to assume the accretion rate varies radially as $\dot{m}_{\rm acc} \propto r^s$, where $0 \leq s \leq1$ (Blandford \& Begelman 1999; Narayan, Igumenshchev \& Abramowicz 2000; Quataert \& Gruzinov 2000; Narayan, Piran \& Kumar 2001; Yuan, Quataert \& Narayan 2003; Kohri, Narayan \& Piran 2005; Begelman 2012).  The case $s=0$ corresponds to the absence of a wind, while $s=1$ implies strong mass loss.  We keep $s$ as a free parameter, which we take to be constant in time.  In the disk's central portions mass accretion is effectively in steady state; therefore  $\dot{m}_{\rm acc}= r^{1/2} \partial (\nu \Sigma r^{1/2})/ \partial r$, implying $\Sigma(r,t) \propto r^{s-n}$.

Strubbe \& Quataert (2009) consider the possibility that an outflow will be launched from $R_f$ due to shock heating during the circularization and when the fallback material joins the disk. Also see Ulmer (1999), and Ayal, Mario \& Piran (2000) for earlier investigations. They parametrize that mass loss -- which exists only when the fallback rate is above the Eddington accretion rate -- with a constant mass loss fraction $\sim 0.1$. This outflow component, if present, will reduce the mass rate that flows toward the black hole within the disk. However, this impact will be effectively absorbed in the disk ejected wind that we just prescribed in the above, so we do not include this outflow component as an additional piece.

Without fallback, the early advective disk spreads under the influence of viscous readjustment modified by wind loss; after a few initial viscous times it tends toward the self-similar expanding state we identify in Appendix \ref{Sec:app-ss}.  Its evolution can be described by the differential equations of global mass and angular momentum conservation (Kumar, Narayan \& Johnson 2008)
\beq    \label{eq:ode-no-fallback}
\frac{d M_d}{d t}  = - {M_d\over t_\nu(r_o) }, ~~~
\frac{d J_d}{d t}  = - F_w {J_d \over t_\nu(r_o)}
\eeq
where $F_w$ is the average ratio of the wind's specific angular momentum to that of the disk.  To simplify matters we ignore the accretion of angular momentum by the black hole; formally, this is valid if the disk innermost radius is $R_i \ll R_o$.  We start at $t=t_{\rm cir}$, when the disk's mass is $M_{\rm ini}$ and its radius is $R_f$. At any time, the disk's angular momentum is $J_d= M_d (GM R_o)^{1/2}$.

Taking $F_w$ to be constant, the solution relevant to an advective disk with Shakura \& Sunyaev viscosity ($n=1/2, q=0$) involves a growing outer disk radius
\beq    \label{eq:ro}
R_o = R_f \left[1+3(1-F_w)\frac{(t-t_{\rm cir})}{t_{\nu0}}\right]^{2/3}.
\eeq
and a decaying disk mass
\beq    \label{eq:Md}
M_d= M_{\rm ini} \left[1+3(1-F_w)\frac{(t-t_{\rm cir})}{t_{\nu0}}\right]^{-\frac{1}{3-3F_w}}.
\eeq
The characteristic normalized accretion rate in the outer disk,  $\dot m_o \equiv \dot{m}_{\rm acc}(r_o)=|\dot M_d|/\dot M_{\rm crit}$, varies as
\beq    \label{eq:mdotacc-adv}
\dot{m}_o= \frac{M_{\rm ini}}{t_{\nu0}\dot M_{\rm crit} } \left[1+3(1-F_w) \frac{(t-t_{\rm cir})}{t_{\nu0}}\right]^{-\eta_{\dot M_d}}
\eeq
with
\beq
\eta_{\dot M_d} = \frac{4-3F_w}{3-3F_w}.
\eeq
Note that, in the presence of later fallback, self-similar expansion is only possible if $|\dot M_d| > \dot M_{\rm fb}$, so that fallback remains negligible.  For this to remain true at late times, one requires $\eta_{\dot M_d} < 5/3$, i.e., $F_w < 1/2$.

The accretion rate at a fixed radius $r<r_o$ is $\dot{m}(r)= \dot{m}_o (r/r_o)^s \propto t^{-\eta}$, with
\beq    \label{eq:eta-of-Fw}
\eta= \frac{1+(3+2s)(1-F_w)}{3(1-F_w)}.
\eeq
The central surface density profile proceeds through a sequence of steady states, so that
$\Sigma(r< r_o) \propto t^{-\eta} r^{s-1/2}$.
An expression for $\eta$, based on a self-similar model in which $Q^-_w(R) \propto Q^+(R)$,  is available in Equation (\ref{eq:SelfSimilarEta}) of Appendix \ref{Sec:app-ss}.  Specifying $n=1/2$ and using this in Equation (\ref{eq:eta-of-Fw}) gives
\beq \label{Fw-from-selfsim}
F_w = {2 s  \over 2s+1}f_j
\eeq
where $f_j$ is the lever-arm, i.e., the factor by which the wind angular momentum exceeds the disk angular momentum at each point in the disk.
For the limit in which the wind angular momentum is not enhanced by a lever arm ($f_j=1$), these solutions reduce to $\eta=4(1+s)/3$ and $F_w = 2s/(2s+1)$, as found by Kumar et al.\ (2008).    Disks are unstable for $F_w>1$ due to the wind-induced instability we discuss in Appendix \ref{Sec:app-ss}.

Once $\dot{m}_o$ declines to the border between regimes (i) and (ii), i.e., Equation (\ref{eq:adv=rad}), the disk state falls off the `ledge' discussed in \S \ref{SS:StateTransitions} directly to the radiatively efficient, gas-pressure dominated regime (iii); as soon as this transition propagates over the entire disk, it becomes radiative and ceases to blow a wind. The time of the transition, $t_{\rm tr}$, satisfies $\dot{m}_o(t_{\rm tr}) \simeq r_o(t_{\rm tr})$, or
\beq \label{eq:t_tr-prime-general}
\frac{(t_{\rm tr}-t_{\rm circ})}{t_f} = {t_{\nu0}\over 3(1-F_w) t_f} \left[ \left(M_{\rm ini} \over r_f t_{\nu0} \dot M_{\rm crit} \right)^{3-3F_w\over 6-5F_w}-1\right].
\eeq

To evaluate these formulae in the context of a TDE requires that we choose the appropriate scales for $M_{\rm ini}$ and $t_{\rm cir}$, which depend on the time required for fallback to circularize.  If circularization is relatively rapid ($n_{\rm cir}\lesssim 1$), it is appropriate to associate this disk with the early fallback, i.e., to replace $M_{\rm ini}$ with $\dot M_f t_f = M_*/5$ and $t_{\rm cir}$ with $t_f$.   On the other hand, if circularization is slow ($n_{\rm cir} \gtrsim 1$), then $M_{\rm ini}$ will grow to $\sim M_*/2$ and $t_{\rm cir}=n_{\rm cir} t_f$.   For our numerical evaluations we will use $n_{\rm cir}=1$.

So, ignoring the small offset $-1$ inside the brackets, we have for the case $F_w= 2/3$, (e.g., when $s=f_j=1$),
\begin{multline}    \label{eq:t_tr-prime-wind}
\frac{(t_{\rm tr}-t_{\rm cir})}{t_f} = 0.75 ~\alpha_{-2}^{-5/8} \beta^{-9/16} (t_f/t_*)^{-1} \\ 
								\times M_6^{-5/8} m_*^{19/16} r_*^{-15/16}.
\end{multline}
The result for the no-wind case $F_w=0$ is similar. For fiducial parameter values, the duration of the initial advective phase is rather short. It can last much longer for slowly evolving disk with higher normalized peak fallback rate, i.e., smaller $\alpha$, $M$, $r_*$, or $t_f/t_*$, or higher $m_*$.

The instantaneous accretion rate of the radiative disk just after $t_{\rm tr}$ is given by the current disk mass $M_d(t_{\rm tr})$ divided by the new viscous time $t_{\nu,{\rm gas}}$. Combining Equations (\ref{eq:tvis-gas}),  (\ref{eq:ro}),  (\ref{eq:Md}),  and (\ref{eq:t_tr-prime-wind}), we find, for $F_w=2/3$,
\beq    \label{eq:mdot-gas-wind}
(\dot{m})_{\rm gas}= 0.54\times10^{-4}~ \alpha_{-2}^{1/8} M_6^{-1.71} (r_*/\beta)^{0.69} m_*^{0.23}
\eeq
and
\beq    \label{eq:tnu-gas-wind}
t_{\nu,\rm gas}= 1.4\times10^5~ \alpha_{-2}^{-1/2} M_6^{5/6}  (r_*/\beta)^{1/4} m_*^{1/12}~ \mbox{yr},
\eeq
while for $F_w= 0$ the results are similar. It is straightforward to see that a higher viscosity ($\alpha$) or smaller disk size (smaller $M$ and higher $\beta$) will give a shorter $t_{\nu,\rm gas}$.

The accretion rate is too low to change the disk mass or radius, until a late stage in which $t \approx t_{\nu, \rm gas}$. Then, so long as there have been no additional perturbations such as gas accretion from the interstellar medium or a new TDE,  the disk enters a new self-similar evolution state in which $R_o \propto t^{3/8}$ and $\eta= 19/16$ (Equation \ref{eq:pringle}), a situation previously considered by Cannizzo et al. (1990), at least until there is a change in the opacity or viscosity.

The above evolution of the initial disk without later fallback is schematically summarized in Figure \ref{fig:no-fallback}.

%%%%%%%%%%%%%%%%%%%%%%%%%%%%%%%%%%%%%%%%%%%%%%%%%%%%%%%%%%%%%%%%%%%%%%%%%%%%%%%%%%%%%%%%%%%%%%%%

\section{Disk evolution with fallback}		\label{sec:with-fallback}

We now consider the evolution of a disk affected by the decline in the fallback at later times.
There are two major evolutionary scenarios, which depend on the degree of inclination between the black hole's spin plane and the orbital plane of the disrupted star.  If these are sufficiently aligned, the stellar fallback stream always intersects the outer disk, and its matter and angular momentum are sure to be deposited near the outer disk boundary.  If instead the disk and hole are sufficiently misaligned then the disk will precess away from the orbital plane (\S \ref{S:Precession}).  This arrangement provides a clear path for the fallback stream to return to the point of disruption (effectively $R_f$).   We shall handle this distinction by assuming that matter arrives at the disk outer radius $R_o$ in the aligned case, but at the much more central radius $R_f$ in the misaligned case.   This is a simplification, for two reasons.  First, even in the misaligned case, the disk and fallback stream will align twice per precession period; and second, the disk's thickness changes as it evolves, so the division between the two regimes is not a fixed angle.   As we shall see below, it is possible for infall to confine the outer disk in the aligned case, whereas this does not occur if new matter arrives only at $R_f$.  In both scenarios there is potential for an expanding disk to strongly affect the central accretion rate, and the existence of a disk-driven wind has a strong influence on whether and when this occurs.

%%%%%%%%%%%%%%%%%%%%%%%%%%%%%%%%%%%%%%%%%%%%%%%%%%%%%%%%%%%%%%%%%%
\subsection{Spin-aligned disruptions} \label{sec:Aligned}

In the spin-aligned scenario the disk remains in the same plane as the returning stellar matter.  Therefore the mass and angular momentum of the fallback stream are incorporated at the outer disk radius $R_o$ rather than the circularization radius $R_f$.   Because this may prevent the disk from growing to large radii, we must account for the influence of fallback on $R_o(t)$.     For this we again follow Kumar et al.\ (2008), who employ an approximate global model to track  the combined influence of accretion and disk-driven winds on a disk formed by stellar collapse.    Kumar et al.\  adopt a single viscous time $t_{\nu}(r_o)$ for all of the disk matter.  This is appropriate for the aligned case, where newly-arriving material arrives at the outer disk, but not for the misaligned case where the viscous time of new matter is $t_\nu(r_f)$, which can be much shorter.    Whereas the specific angular momentum of newly-arriving matter increases with time in the collapsar context considered by Kumar et al., in TDEs it remains fixed at $j_{\rm fb} = (GMR_f)^{1/2}$.   This leads to a significant difference in behavior, as we shall see.

The disk differential equations for mass and angular momentum conservation are same as Equation (\ref{eq:ode-no-fallback}) except that fallback terms now appear:
\beq \label{eq:M-ode-aligned}
\left\{
\begin{split}
\frac{d M_d}{d t} &= \dot M_{\rm fb} - {M_d\over t_\nu(r_o) }, \\
\frac{d J_d}{d t} &= j_{\rm fb} \dot M_{\rm fb} - F_w {J_d \over t_\nu(r_o)}.
\end{split}
\right.
\eeq
If we make the restriction that the disk is advective throughout, then it follows that $F_w$ is constant and that $t_\nu(r_o) \simeq J_d^3/(\alpha G^2 M^2 M_d^3)$.   While the disk remains advective, the solution to Equations (\ref{eq:M-ode-aligned}) is therefore governed by the two dimensionless parameters $F_w$ and $t_{\nu0}/t_f$ as well as the dimensional parameters $j_{\rm fb}$, $GM$, and $\dot M_f$.

A disk described by these equations can exist in, and transition between, three asymptotic states:

\noindent -- {\em Transient}: There has been no time for viscosity to act, so no mass or angular momentum has been shed: the accretion terms on the right-hand sides of these equations are negligible.  Accordingly, the disk radius equals the circularization radius $R_f$.  However, this phase cannot last longer than a single viscous time at $R_f$.  The initial, advective disk is in this transient phase for the short period $t_{\nu0}$.

\noindent -- {\em Self-similar spreading}: fallback is negligible and viscous accretion balances the time derivatives on the left-hand side.     The evolution is therefore identical to what we found for intermediate times $t_f <t<t_{\rm tr}$ and late times $t > t_{\nu, {\rm gas}}$ in the no-fallback case considered in \S \ref{sec:no-fallback}.    This state is only accessible if the fallback term in each equation becomes increasingly negligible over time; for it to persist to very late times (when $\dot M_{\rm fb}\propto t^{-5/3}$ but definitely before $t_{\rm tr}$),  it requires $F_w<1/2$.   Nevertheless there can be an extended period of expansion even for larger values of $F_w$, as we shall see.

\noindent -- {\em Steady state}: Newly-incorporated matter is processed rapidly and the terms on the right-hand side effectively cancel.  This  requires $F_w J_d = M_d j_{\rm fb}$, so that $R_o = F_w^{-2} R_f$. To arrive in this state, the disk radius either expands by a factor $F_w^{-2}$ from its value in the transient state, or contracts from the previous self-similar expanding state. 

%%%%%%%%%%%%%%%%%%%%%%%%%%%%%%%%%%%%%%%%%%%%%%%%%%%%%%%%%%%%%%%%%%%%%%%%%%%%%%%
\begin{figure}
\centerline{
\includegraphics[width=8.5cm, angle=0]{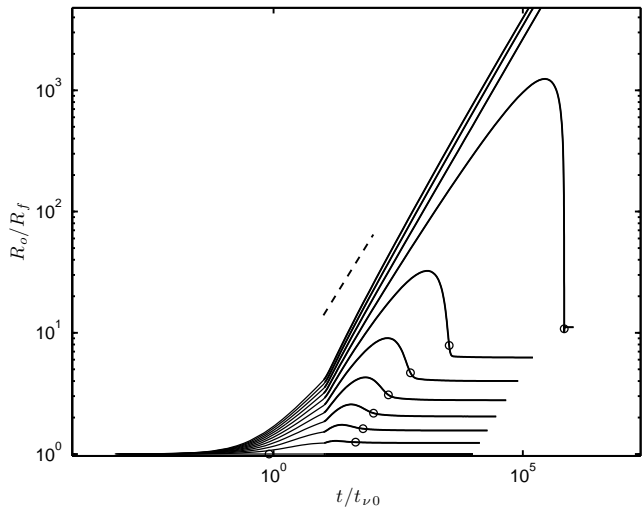}
}
\caption{Evolution of advective, aligned disks, with $F_w=0$ (top curve) to $F_w=1$ (bottom curve) in steps of 0.1.   In these examples, the fallback rate is constant before $t_f = 10 t_{\nu0}$, then drops $\propto t^{-5/3}$.  A transient phase persists while $t\ll t_{\nu0}$, during which $R_o = R_f$.  Afterwards the disk tends either towards the steady-state value $R_o=F_w^{-2} R_f$ or toward a self-similar expanding state $R_o\propto t^{2/3}$ (dashed line).  For  those curves which expand beyond the steady-state radius before contracting again, we assign $t_{\rm contr}$ (circles) to be the time of maximum contraction. In making this figure we ignore the circularization process and evolve the disk from $t=0$ to deliberately show the transient phase behavior.}       \label{fig:contract-examples}
\end{figure}
%%%%%%%%%%%%%%%%%%%%%%%%%%%%%%%%%%%%%%%%%%%%%%%%%%%%%%%%%%%%%%%%%%%%%%%%%%%

%%%%%%%%%%%%%%%%%%%%%%%%%%%%%%%%%%%%%%%%%%%%%%%%%%%%%%%%%%%%%%%%%%%%%%%%%%%%%%%
\begin{figure}
\centerline{
\includegraphics[width=8.5cm, angle=0]{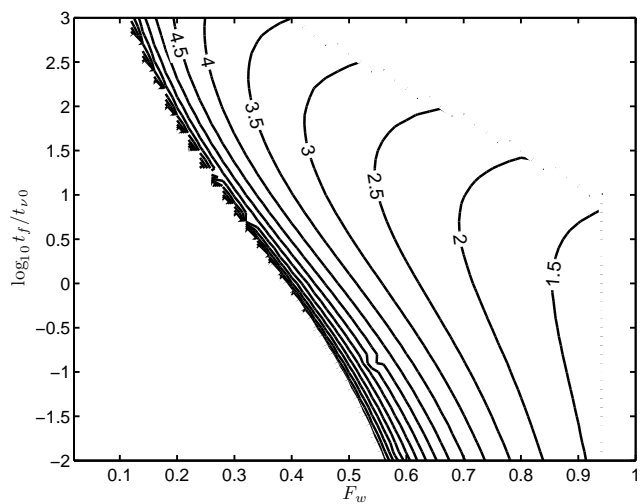}
}
\caption{Contours of $\log_{10}t_{\rm contr}/t_{\nu0}$ for nonradiative aligned TDE disks, where $t_{\rm contr}$ is the time of maximum contraction discussed in  figure \ref{fig:contract-examples}.}		\label{fig:contract-map}
\end{figure}
%%%%%%%%%%%%%%%%%%%%%%%%%%%%%%%%%%%%%%%%%%%%%%%%%%%%%%%%%%%%%%%%%%%%%%%%%%%

Although Equation (\ref{eq:M-ode-aligned}) applies to both advective and radiative aligned disks, it is important to realize that there is no `steady state' solution in the absence of a wind ($F_w=0$).  Therefore radiative disks can only be `transient', when they are younger than one radiative viscous time, or `self-similarly spreading', once viscosity becomes important.

Advective disks described by Equation (\ref{eq:M-ode-aligned}) do not necessarily persist in either the steady or self-similar state after the initial transient period.  At the beginning of the accretion, the disk typically enters a phase of self-similar expansion.  However, depending on the values of $F_w$ and $t_f/t_{\nu0}$, its radius may decline rapidly at some later time $t_{\rm contr}$, before leveling out at the steady state radius $F_w^{-2} R_f$.  We show examples of the evolution of advective disks in Figure \ref{fig:contract-examples}.

This contraction behavior occurs when both fallback and winds are important in the disk evolution.  It does not occur in the absence of fallback, as we saw in \S \ref{sec:no-fallback}.  Nor does it occur for sufficiently weak winds, as is clear in Figures \ref{fig:contract-examples} and \ref{fig:contract-map}.

Bear in mind that an advective disk will become radiative at some point, so for some cases the disk will still be expanding when it transitions to a radiative state.   This can be seen in Figure \ref{fig:contract-map}, in which we display the parameter dependence of $t_{\rm contr}$ for advective disks.   Because fallback terms are relatively minor in the expanding state, the presence of fallback adds only a small delay to the advective-radiative transition time given by Equation (\ref{eq:t_tr-prime-wind}).

\begin{figure*}
\begin{center}
\includegraphics[width=8.8cm, height= 6.5cm, angle=0]{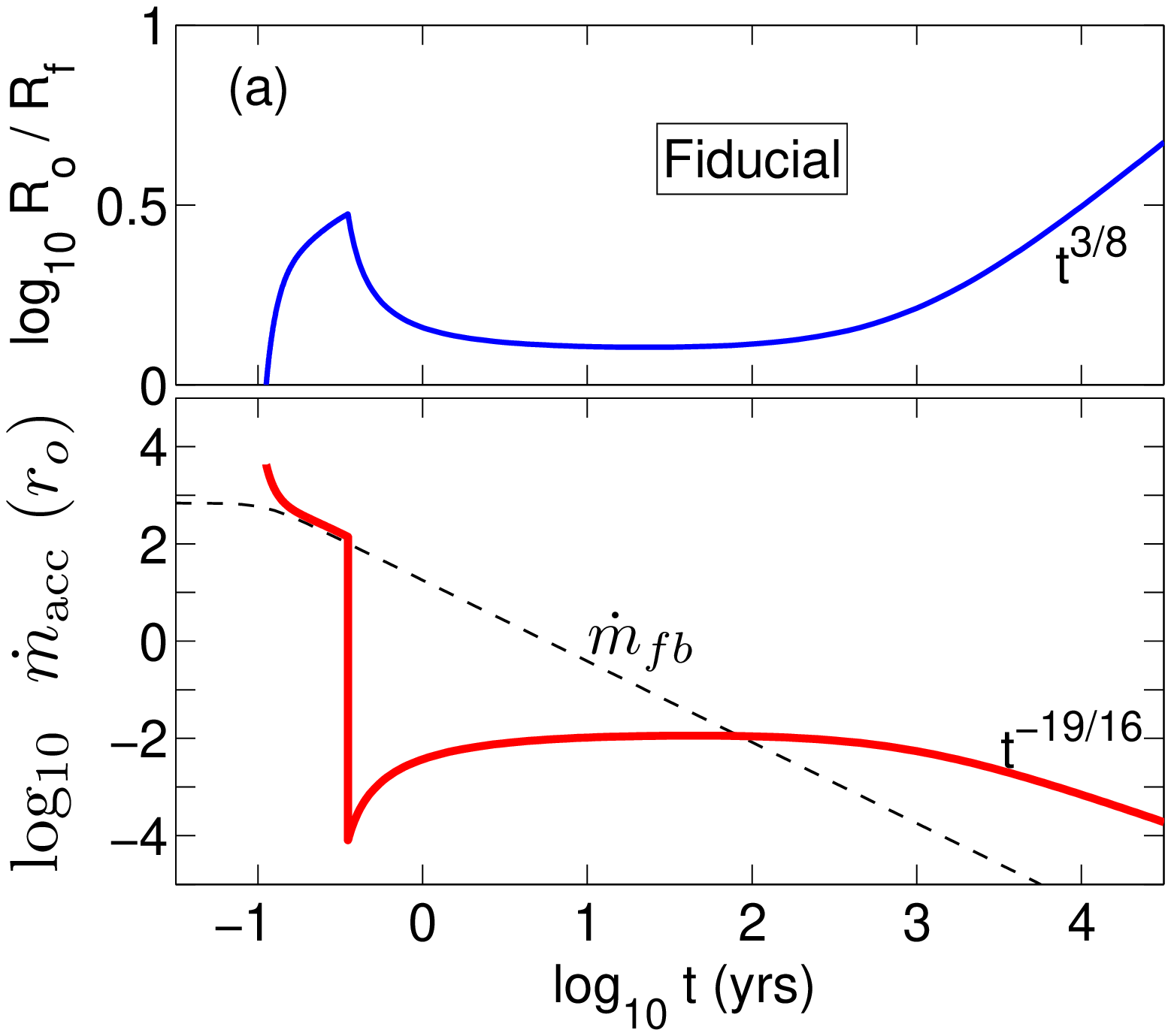}
\includegraphics[width=8.8cm, height= 6.5cm, angle=0]{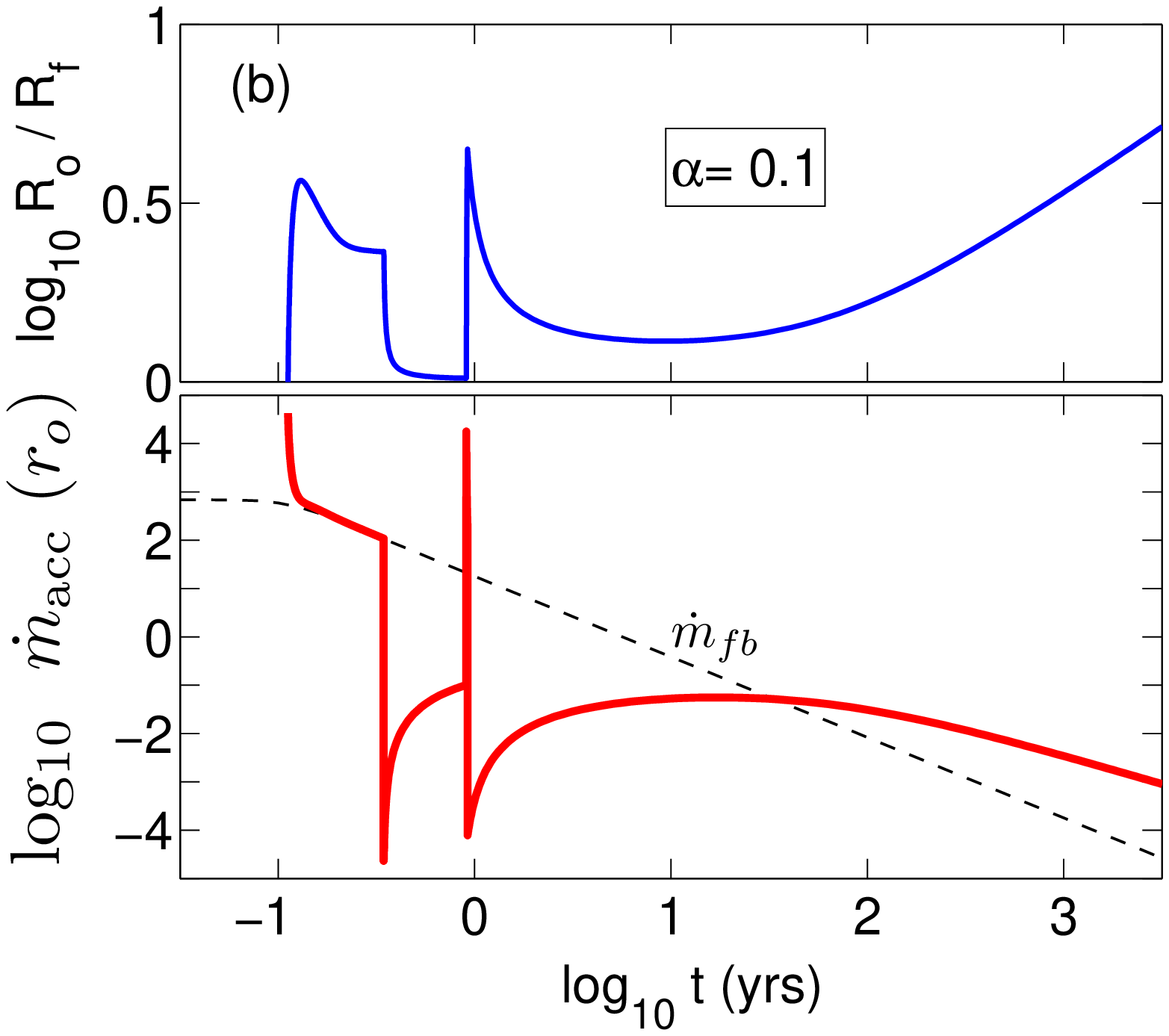}
\end{center}
\begin{center}
\includegraphics[width=8.8cm, height= 6.5cm, angle=0]{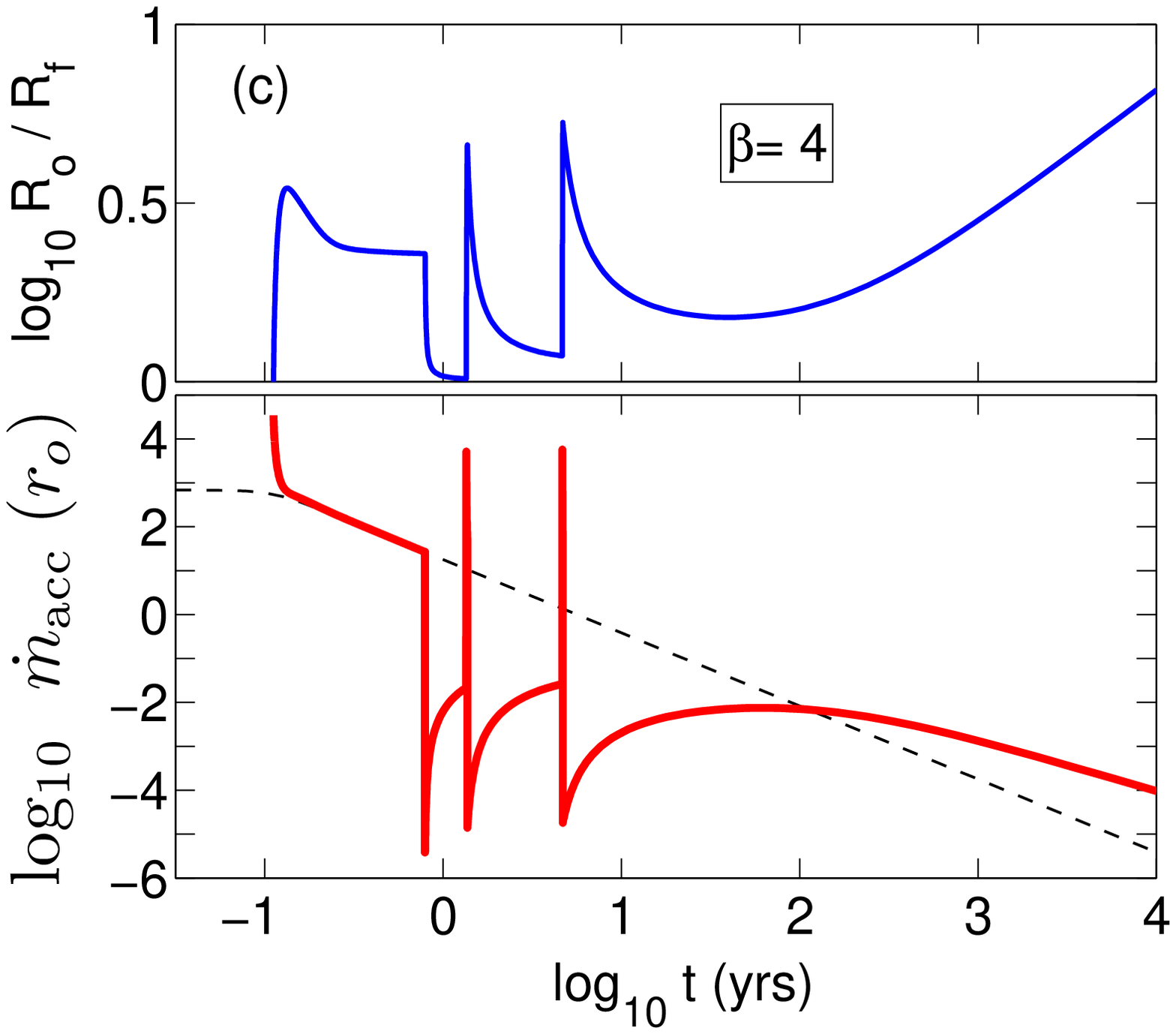}
\includegraphics[width=8.8cm, height= 6.5cm, angle=0]{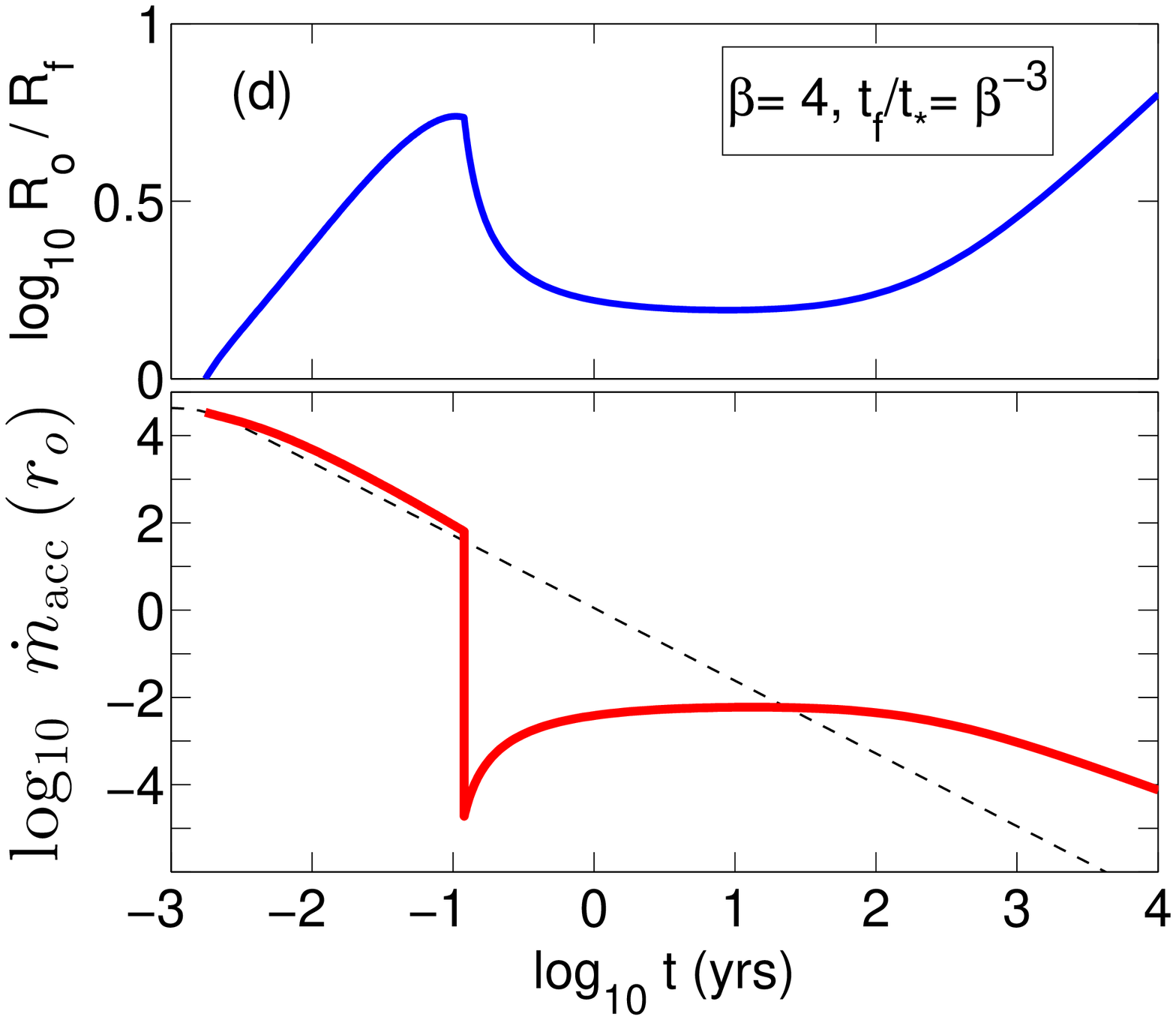}
\end{center}
\begin{center}
\includegraphics[width=8.8cm, height= 6.5cm, angle=0]{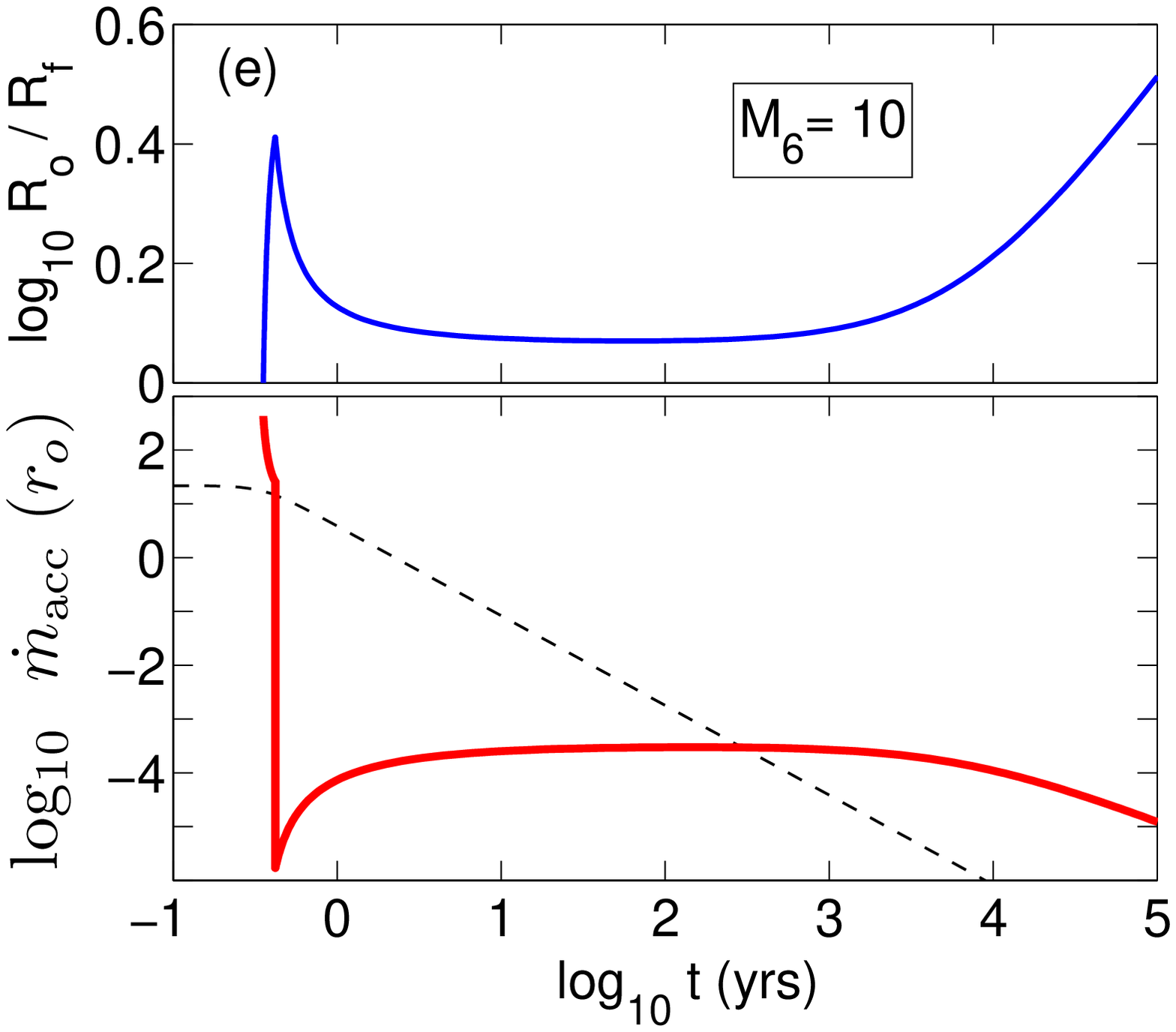}
\includegraphics[width=8.8cm, height= 6.5cm, angle=0]{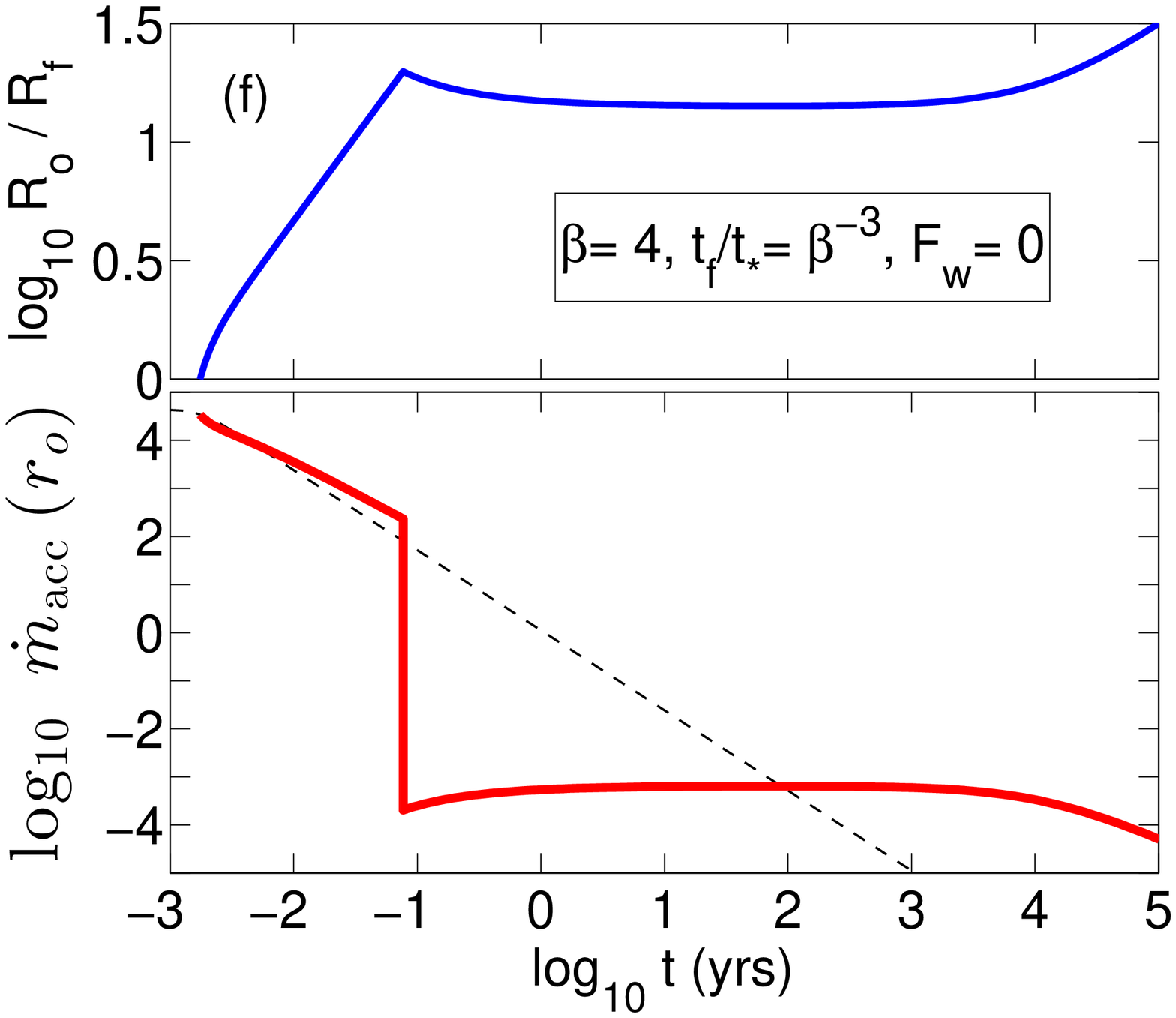}
\end{center}

\caption{The evolution of the disk with fallback aligned with the black hole equator plane, calculated from Equation (\ref{eq:M-ode-aligned}) for varying model parameter values. For each panel, the top sub-panel is for the disk outer radius $R_o$, and the bottom sub-panel is for the accretion rate at $R_o$. The dashed line is the fallback rate. Panel (a) is for the fiducial model parameter values ($\beta$, $t_f/t_*$, $m_*$, $r_*$, $M_6$, $\alpha_{-2}$, $n_{\rm cir}$) $= 1$ and $F_w=2/3$. In subsequent panels the parameter(s) that vary from the fiducial values are labeled. In some panels, the initial high peak of accretion rate at $t_f$ is contributed from the accretion of the initial disk whose mass is accumulated during the circularization $t<t_{\rm cir}$.}		 \label{fig:with-fallback-aligned}
\end{figure*}

The presence of fallback, which adds low-angular-momentum matter to the outer disk, causes the disk to undergo oscillations once it becomes radiative.  This behavior is related to the limit cycles discussed in \S \ref{SS:StateTransitions}, except that it involves changes in the disk's radius as well as its thermal state.
The oscillation is shown in two cases in Figure \ref{fig:with-fallback-aligned}
in which we evolve Equation (\ref{eq:M-ode-aligned}) over time for a range of model parameters.  Once the disk crosses the `ledge' $(\dot{m}_{\rm acc})_{\rm i-ii}$, its temperature drops from its value in the advective state (i) down to that in the radiative state (iii). Instead of then gradually draining over its new viscous time, the disk now accumulates matter through fallback.  The addition of low angular momentum material causes the disk to shrink and its surface density to increase.  If the disk can acquire more mass from fallback in a single $t_{\nu,{\rm gas}}$ than it had in the advective stage, i.e., if $M_d(t) < \int_t^{t+t_{\nu,{\rm gas}}} \dot M_{\rm fb}(t') dt'$, then the disk will shrink dramatically as it is pushed back toward the `transient' state: $R_o\rightarrow R_f$, as is shown in panels (a -- e) of Figure \ref{fig:with-fallback-aligned}. We advise that this rapid shrinking, which occurs at the advective-radiative transition, not be confused with the contraction described earlier, which happens only in the advective phase.

The disk cannot reach $R_f$ in the first cycle if it was expanding prior to becoming radiative, because self-similar expansion of an aligned disk requires $|\dot M_d|>\dot M_{\rm fb}$.  This implies that $M_d$ exceeds $(3/2) t \dot M_{\rm fb}(t)$, the total mass of future fallback, so the disk mass cannot increase much.    This is expected, for instance, in the case where there is no disk wind at all ($F_w=0$), as such disks did not experience contraction in the advective phase [see panel (f) of Figure \ref{fig:with-fallback-aligned}].

On the other hand, if the disk experienced a contraction prior to the advective-radiative transition, then its radius will already have contracted to $F_w^{-2} R_f$ and its mass will have dropped to its correspondingly low steady-state value $F_w^{-1} t_{\nu0} \dot M_{\rm fb}(t)$.   In that case, fallback accumulation onto the radiative disk does overwhelm its initial mass, pushing the radius down to $R_f$.  Two examples of the latter scenario are given in Figure \ref{fig:with-fallback-aligned}, one for $\alpha=0.1$ and the other for $\beta=4$ (panels b and c). Both cases have smaller $t_{\nu0}/t_f$ (Equation \ref{eq:tnu0}) so to leave enough time for disk to contract before becoming radiative.

The increase of $\Sigma$ and the decrease of $R_o$ bring the radiative disk to the border between regimes (iii) and (ii),  $(\dot{m}_{\rm acc})_{\rm ii-iii}$.
The disk temperature then jumps back up to its advective value, triggering a rapid accretion of mass and a expansion of the disk radius on the advective viscous time scale.   The expansion proceeds self-similarly until $\dot M_d$ matches the current rate of infall $\dot M_{\rm fb}(t)$; at this point the disk radius stops expanding and moves toward the steady-state value $F_w^{-2} R_f$.   However a steady state cannot be achieved while $\dot M_{\rm fb}$ is within the thermally-unstable range of accretion rates at this radius; the disk must again become radiative, repeating the cycle.   After the first cycle the disk is effectively drained of its original mass; the advective phase then lasts only a few times $t_{\nu0}$, and all the properties of the cycle are determined by the current fallback rate $\dot M_{\rm fb}$.

%%%%%%%%%%%%%%%%%%%%%%%%%%%%%%%%%%%%%%%%%%%%%%%%%%%%%%%%%%%%%%%%%%%%%%%%%%%%%%%
%%%%%%%%%%%%%%%%%%%%%%%%%%%%%%%%%%%%%%%%%%%%%%%%%%%%%%%%%%%%%%%%%%%%%%%%%%%

The timing of the first advective-to-radiative transition depends on whether the advective disk's radius is still expanding or has contracted at the transition time.  If it is still expanding, the transition time $t_{\rm tr}$ is close to the value we identified for a self-similar disk without fallback; see Eqs. (\ref{eq:t_tr-prime-general}) and (\ref{eq:t_tr-prime-wind})  (there is a minor delay relative to those estimates, caused by the addition of fallback).  On the other hand, if the transition occurs after $t_{\rm contr}$, then $R_o = F_w^{-2} R_f$, the accretion rate at $R_o$ equals $\dot M_{\rm fb}(t)$,  and the criterion for transition becomes  $\dot{m}_{\rm fb}(t_{\rm tr}) \simeq r_f/F_w^2$:
\bea	\label{eq:t_tr-post-contr}
\frac{t_{\rm tr}}{t_f} &=& \left(\frac{F_w^2 \dot{m}_f}{r_f}\right)^{3/5} \\
 &=& 
3.1 \left(F_w\over2/3\right)^{6/5} \left( \frac{\beta}{t_f/t_*} \right)^{3/5} M_6^{-1/2} r_*^{-3/2} m_*^{7/5}. \nonumber
\eea
This is typically later than the transition time of a still-expanding disk.  If it is earlier (which can happen for low values of $\alpha$ and $k$),  then it becomes possible for the contraction itself to  stimulate the transition.

The duration of the cycle is dominated by the accumulation of fallback material onto the radiative disk, which pushes it up across the radiative-advective boundary; this occurs when $\dot{m}_{\rm acc}= M_d/(t_{\nu, \rm gas} \dot{M}_{\rm crit})= (\dot m_{\rm acc})_{\rm ii-iii}$ at the outer disk radius, $R=R_o$.   In most cases (except for the first radiative period after a phase of advective expansion), $R_o$ is pushed down to $R_f$ by the arrival of material, so we can apply this criterion at $R=R_f$. A radiative disk of radius $R_f$ becomes advective when its mass reaches the critical value
\beq 	\label{eq:Mcrit:rad->adv}
M_{d,{\rm crit}} = t_{\nu, \rm gas}(r_f) (\dot m_{\rm acc})_{\rm ii-iii} \dot{M}_{\rm crit}
\eeq
Moreover, only a disk which begins the radiative phase with negligible mass has $R_o=R_f$.  The duration of the radiative phase is therefore set by the time required to accumulate $M_{d,{\rm crit}}$: integrating $\dot M_{\rm fb}$ over time, this implies that the change in $(t_f/t)^{2/3}$ during an entire radiative phase is given by
\beq 	\label{eq:RadPhaseDuration-General}
\delta \left[\left(t_f\over t\right)^{2/3}\right]_{\rm rad} = \left(t_f\over t_1\right)^{2/3}
\eeq
where 
\beq 	\label{eq:t1}
\begin{split}
{t_1 \over t_f } &= \left(\frac32 {\dot M_f t_f \over M_{d,{\rm crit}}}\right)^{3/2} \\
&= 0.48\alpha_{-2}^{1.31}  (\beta/r_*)^{3.28} m_*^{2.59}
\end{split}
\eeq
is the time after which the total mass of future fallback is less $M_{d,{\rm crit}}$; therefore advective cycles are no longer possible in disks which become radiative after $t_1$. It is possible to have no cycles at all, because $t_{\rm tr} > t_1$; indeed, this is will be the case for the fiducial parameter values (comparing Equations \ref{eq:t_tr-post-contr} and \ref{eq:t1}), and four panels of Figure \ref{fig:with-fallback-aligned} show additional examples of this. Larger $\alpha$ or $\beta$ corresponds to later $t_1$, i.e., easier to have cycles, because these correspond to smaller $t_{\nu,\rm gas}(r_f)$ and $(\dot m_{\rm acc})_{\rm ii-iii}$ thus lower $M_{d,{\rm crit}}$.

For the radiative phase occurrence time $t\ll t_1$, the duration of the radiative phase is given by
\beq	\label{eq:delta-t_rad-approx}
\frac{(\delta t)_{\rm rad}}{t} \simeq \frac{3}{2}\left(\frac{t}{t_1}\right)^{2/3},
\eeq
i.e., when a sequence of advective cycles exist, the duration of each preceding radiative phase (to accumulate fallback mass) becomes increasingly longer; this is seen in panel (c) of Figure \ref{fig:with-fallback-aligned}.

It is interesting to note that thermal cycles end because fallback no longer supplies sufficient mass to trigger them, not because the fallback rate falls to the rate that can be processed stably by a radiative disk [i.e., not $(\dot m_{\rm acc})_{\rm ii-iii}$].  This is a consequence of the fact that the radiative disk has a viscous time much longer than $t_1$.  

The advective pulse of each cycle is identical, because each one starts with a disk of radius $R_f$, mass $M_{d,{\rm crit}}$, and negligible fallback within the advective viscous time, $\dot M_{\rm fb}(t) t_{\nu0}\ll M_{d,{\rm crit}}$.  (The first cycle after a long self-similar advective phase is an exception, as noted above, because disks in this case did not experience a contraction before $t_{\rm tr}$, thus do not shrink all the way to $R_f$ after becoming radiative.)   The peak accretion rate is $\dot M_{\rm pulse} = M_{d,{\rm crit}}/t_{\nu0}$, or
\beq	\label{eq:mdot_pulse}
\dot m_{\rm pulse} = 1.3\times10^4~\alpha_{-2}^{1/8} M_6^{-1/3} (r_*/\beta)^{11/16} m_*^{-0.23}.
\eeq
The duration of each advective pulse is determined by precisely the same dynamics which led to Equation (\ref{eq:t_tr-prime-general}), except that the disk mass is initially $M_{d,{\rm crit}}$ rather than $\dot M_f t_f$.   The advective pulse lasts $\sim 6.5 t_{\nu0}$ in the case $F_w=0$, and $\sim 1.7 t_{\nu0}$ for $F_w=2/3$. The duty cycles of these pulses, if they exist, are low: e.g., $\sim t_{\nu0}/(\delta t)_{\rm rad} < 10^{-3}$ for $\beta= 4$. Thus, the pulses may be too brief to be observable.  

After its radiative transition, the disk accumulates low-angular-momentum matter, causing its radius to shrink toward $R_f$. Unless the prior advective period involved a long phase of self-similar expansion ($F_w\ll1$), this process completes and the disk re-enters an `transient' phase with $R_o=R_f$.  Because this is only possible if matter accumulates quickly compared to the viscous time,
the instantaneous accretion rate rises to a limiting value equal to the mass accumulated since $t_{\rm tr}$ divided by  $t_{\nu, \rm gas}(r_f)$. The accumulated mass is $\int_{t_{\rm tr}}^{\infty} \dot{M}_{\rm fb}(t) dt= (3/2)t_{\rm tr} r_f \dot{M}_{\rm crit}/F_w^2$, where we have used $\dot{m}_{\rm fb}(t_{\rm tr}) \simeq r_f/F_w^2$, and the viscous time is given by Equation (\ref{eq:tvis-gas}).  So, if there are no further advective cycles, the radiative-phase accretion rate approaches 
\begin{multline} \label{mdot_gas,nocycle}
(\dot{m})_{\rm gas, no\,cycle} = 0.0097~ \alpha_{-2}^{4/3} \beta^{5/3} (t_f/t_*)^{2/3} F_w^{-4/3}\\ 
								\times M_6^{-14/9} m_*^{8/9} r_*^{-2/3}.
\end{multline}
However, when this value exceeds the maximum allowed in the radiative, gas-pressure dominated regime (cf. Equation \ref{eq:rad=gas}),
\beq
(\dot{m})_{\rm ii-iii}(r_f)= 0.13~ \alpha_{-2}^{-1/8} M_6^{-1} (r_*/\beta)^{21/16} m_*^{-7/16},
\eeq
the disk transitions back into the advective regime, depletes in mass quickly within a time $\sim t_{\nu0}$, before landing in the radiative regime again -- thus rendering a cycle.  In this case, we can use $(\dot{m})_{\rm ii-iii}(r_f)$ as an approximation to the accretion rate at the time when there is no future cycle and before the next evolution stage begins. This approximation becomes closer as the number of cycles that the disk experienced increases. 
Combined with Equation (\ref{mdot_gas,nocycle}), the true estimate of the accretion rate right before $t_{\nu, \rm gas}$ is the minimum of the two:
\beq    \label{eq:mdot-gas-true}
(\dot{m})_{\rm gas} \simeq \min[(\dot{m})_{\rm gas, no\,cycle}, (\dot{m})_{\rm ii-iii}(r_f)],
\eeq
where the two values are relevant in the absence or presence of advective cycles, respectively.

The time when the disk enters the self-similar state of the gas-pressure dominated regime, $t_{\nu, \rm gas}$, is given by Equation (\ref{eq:tvis-gas}). The accretion rate there is from Equation (\ref{eq:mdot-gas-true}) and the radius is $r_f$. Thus,
\bea
t_{\nu, \rm gas} & \simeq & \max [ 
~4100~ \alpha_{-2}^{-4/3} \beta^{-31/15} (t_f/t_*)^{-4/15} F_w^{8/15} \nonumber\\ 
				&  & ~~~~~~~\times M_6^{8/9} r_*^{5/3} m_*^{-37/45},\\
				&  & ~~~~~~~ 1400~ \alpha_{-2}^{-3/4} M_6^{2/3} (r_*/\beta)^{7/8}  m_*^{-7/24} ] ~\mbox{yrs} \nonumber.
\eea

%%%%%%%%%%%%%%%%%%%%%%%%%%%%%%%%%%%%%%%%%%%%%%%%%%%%%%%%%%%%%%%%%%%%%%%%%%%%%%%%%%%%%%%%%%%%%%%%%

\subsection{Spin-misaligned disruptions}	\label{subsec:Misaligned}

We treat the case of a disk misaligned with the spin plane of its central black hole by assuming that the Lense-Thirring precession causes it to precess away from the plane of the original stellar orbit (\S~\ref{S:Precession}).  Then, for most of the time (except when the disk realigns with it), the infall stream misses the outer disk and has a clear path to the original stellar pericenter. We therefore consider separately the outer, expanding relic of early accretion, and the inner disk which receives matter from the fallback stream as well as the outer disk.   We do not use Equation (\ref{eq:M-ode-aligned}) to treat the fallback, because the viscous time at $R_f$ is much less than that of the entire disk.

During its advective phase, the disk viscosity is independent of surface density. Under the assumptions we adopt in Appendix \ref{Sec:app-ss}, the evolution equation (\ref{eq:DiskEvlnWithWind}) is linear in $\Sigma$ and can be solved with Green's functions even when a wind is present.  This appears to be a novel point, as we only know of Green's function solutions for wind-free disks.   In Appendix \ref{sec:app-green} we use Green's functions to examine the properties of an advective disk which spreads while matter is added at $R_f$. We restrict that analysis to the wind-free case, but several lessons can be generalized to the windy case.

\begin{figure}
\centerline{
\includegraphics[width=8.8cm, angle=0]{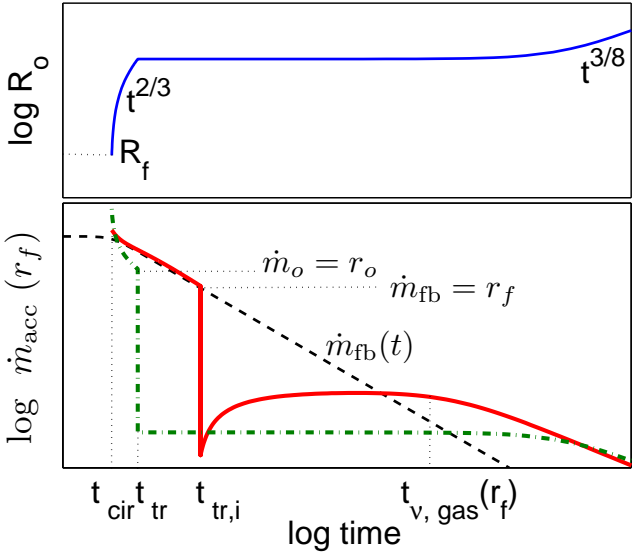}
}
\caption{Schematic evolution of the TDE disk that is misaligned with the black hole spin plane. \textit{Top} panel shows the outer disk radius. \textit{Bottom} panel shows the accretion rate at $R_f$ (red solid line). The green dot-dashed line represents the contribution to $\dot{m}_{\rm acc}(r_f)$ from the spreading initial disk, and it declines as $t^{-\eta}$ in the advective phase. In this figure we consider the strongly windy disk case such that $\eta > 5/3$, thus $\dot{m}_{\rm acc}(r_f)$ always balances the mass supply rate from fallback, therefore tracks $\dot{m}_{\rm fb}(t)$, until the inner disk transitions to the regime (iii) at $t_{\rm tr,i}$. The spreading outer disk has transitioned to regime (iii) at an earlier time $t_{\rm tr}$. If the advective disk were weakly windy such that $\eta < 5/3$, then $\dot{m}_{\rm acc}(r_f)$ would initially fall shallowly as $t^{-\eta}$, then suddenly drop to $\dot{m}_{\rm fb}(t)$ at $t_{\rm tr}$; the behavior afterwards would be same as depicted here.}    \label{fig:misaligned}
\end{figure}

First, the outer disk is essentially unaffected by the addition of matter at $R_f$, because the remnant of early accretion always expands self-similarly (Appendix \ref{Sec:app-ss}) beyond what arrives later, and because the the disk is the sum of the two contributions.   Second, the rate of mass accretion onto the black hole is a superposition of viscous accretion from the outer disk and accretion driven by current fallback.  Therefore, if the outer disk has $\eta <5/3$, i.e., if it has $s<1/4$ for $f_j=1$ (Figure \ref{fig:selfsimdisk}) its contribution will always dominate the central flow at late times, implying $\dot m_{\rm acc}(r_i)\propto t^{-\eta}$; otherwise, fallback dominates central accretion and $\dot m_{\rm acc}(r_i)\propto t^{-5/3}$. The latter case is illustrated in Figure \ref{fig:misaligned}.  Third, the additional surface density created by newly-incorporated matter is proportional to $R^{s-n}$ at radii smaller than $R_f$, but steepens to $R^{-K/3-n}$ for radii between $R_f$ and $R_o$, where $K$ is defined for a windy disk in Equation (\ref{eq:K-for-wind}).

Because the outer disk is expanding self-similarly, its transition from the advective to radiative state occurs at the time $t_{\rm tr}$ we identified in Equations (\ref{eq:t_tr-prime-general})-(\ref{eq:t_tr-prime-wind}).  The transition then works its way inward within a single viscous time.  If the evolution law of the advective outer disk was such that $\eta <5/3$, then it was up to this point the dominant source of accretion for the black hole.  In this case there is a sudden drop of the central accretion rate to the current fallback rate, $\dot M_{\rm fb}(t)$.

One might expect that outer disk's transition to a radiative state would trigger the central disk to become radiative as well, but the central disk is fed directly by fallback.  It therefore cannot become radiative until $t_{\rm tr,i}$, which is when $\dot m_{\rm fb} \simeq r_f$:
\beq \label{eq:t_tr_innerdisk}
\begin{split}
t_{\rm tr,i} &= (\dot{m}_f /r_f)^{3/5} t_f \\
  & =  205\, \beta^{3/5} (t_f/t_*)^{2/5} m_*^{2/5} \,{\rm days}.
\end{split}
\eeq
In general, the transition time for the inner disk is the later of $t_{\rm tr}$ and $t_{\rm tr,i}$ (see Figure \ref{fig:misaligned}).

In principle the transition of the inner disk from an advective to a radiative state can be followed by thermal pulses, as we predicted for aligned disks in \S \ref{sec:Aligned}.   In fact, however, this may not happen: comparing $t_{\rm tr, i}$ to the critical time $t_1$ (Equation \ref{eq:t1}),
\beq\label{eq:t_tr_i-compared-to-t_1}
{t_{\rm tr,i} \over t_1} = 10~ \alpha_{-2}^{-1.31} \beta^{-2.68} (t_f/t_*)^{-3/5} M_6^{1/2} r_*^{1.78} m_*^{-1.19}.
\eeq
For our fiducial parameters, the central disk will never receive enough fallback to stimulate an advective pulse.

Once the inner disk has entered the radiative phase, accretion onto the black hole is determined by the viscous evolution of a radiative, gas-pressure dominated disk.
For a misaligned disk, our theory implies that there are in fact two separate mass reservoirs for this late-time accretion.  One is the outer disk, whose mass accretion rate equals $(\dot m)_{\rm gas}$ listed previously in Equation (\ref{eq:mdot-gas-wind}).  After a single outer viscous time [$t_{\nu,{\rm gas}}(r_o)\sim 10^{5.3}$ years: Equation (\ref{eq:tnu-gas-wind})], it enters self-similar spreading with $R_o\propto t^{3/8}$ and $\dot m_{\rm acc} \propto t^{-19/16}$.

A second mass reservoir is the inner disk at $R_f$, which has acquired new matter from the fallback stream.  Assuming it accumulates all the fallback from $t_{\rm tr,i}$ so that its mass is $(3/2) t_{\rm tr,i} \dot M_{\rm fb} (t_{\rm tr,i})$, its viscous time is solved from Equation (\ref{eq:tvis-gas}) to be
\begin{multline} \label{eq:t_nu_inner_radiative_disk}
t_{\nu,{\rm gas}}(r_f) = 4100~ \alpha_{-2}^{-4/3} \beta^{-31/15} (t_f/t_*)^{-4/15}\\ 
				            \times M_6^{8/9} r_*^{5/3} m_*^{-0.82}~{\rm yrs}
\end{multline}
and its dimensionless accretion rate is
\begin{multline} \label{eq:mdot_inner_radiative_disk}
(\dot m)_{\rm gas}(r_f) = 0.0097~ \alpha_{-2}^{4/3} \beta^{5/3} (t_f/t_*)^{2/3}\\
 						\times M_6^{-14/9} m_*^{8/9} r_*^{-2/3}.
\end{multline}

Contributions of both reservoirs are shown in Figure \ref{fig:misaligned}.
Because the inner disk has a higher accretion rate than the outer disk, it is guaranteed to dominate black hole accretion for several inner viscous times, or tens of thousands of years, in the absence of any other perturbations.

%%%%%%%%%%%%%%%%%%%%%%%%%%%%%%%%%%%%%%%%%%%%%%%%%%%%%%%%%%%%%%%%%%%%%%%%%%%%%%%
%%%%%%%%%%%%%%%%%%%%%%%%%%%%%%%%%%%%%%%%%%%%%%%%%%%%%%%%%%%%%%%%%%%%%%%%%%%

%%%%%%%%%%%%%%%%%%%%%%%%%%%%%%%%%%%%%%%%%%%%%%%%%%%%%%%%%%%%%%%%%%%%%%%%%%%%%%%%%%%%%
\section{Disk precession and its evolution}\label{S:Precession}

Precession of the TDE disk could modulate the light curve in several ways.  In addition to presenting a variable disk orientation to the observer,  it is likely that a disk wind would interfere with jet emission from the central source in a way that changes periodically as the disk precesses, possibly by deflecting the jet. It is even possible that periodic interruptions of the fallback stream would imprint themselves on the brightness evolution.  If such a signal is observed, it is most likely to be due to frame dragging by a spinning black hole inclined to the orbital plane of the disrupted star.

For a test particle that is in a circular orbit around a BH of mass $M$ but whose orbital plane is misaligned with the
central object's equatorial plane, the general relativistic dragging of inertial frames
 causes the particle's orbital plane precess at an angular speed  ${\mathbf \Omega}_{\rm LT} = 2{\mathbf L}/R^3$,
(Bardeen \& Petterson 1975; Ciufolini et al. 1998) with units $G=c=1$, where $L= a M^2$ is the BH angular momentum, $a$ is the dimensionless BH spin parameter, and $R$ is the orbit radius.   This is the Lense-Thirring effect (Lense \& Thirring 1918; Mashhoon et al. 1984), equivalent to a torque ${\mathbf \tau}= {\mathbf \Omega}_{\rm LT} \times {\mathbf J}$ acting on the orbital angular momentum ${\mathbf J}$.

Because of its strong radial dependence, frame dragging acts most rapidly on the inner regions of the disk.   Its effect depends on the propagation rate of a disk warp relative to the viscous inflow rate $v_r=R/t_\nu$ and the local orbital precession rate $\Omega_{LT}$.   Warps propagate either diffusively with a diffusivity $\sim \nu/(2\alpha^2)$ (Papoloiziou \& Pringle 1983) or as waves (Pringle 1999; Nelson \& Papaloizou 1999), so that the propagation speed over a scale $R$ is
\beq\label{eq:v_warp-warp-velocity}
v_{\rm warp} \simeq \min\left({c_s\over 2}, {\nu \over 2 \alpha^2 R}\right),
\eeq
assuming the vertical shear viscosity is comparable to $\nu$.  Propagation is wavelike in the low-viscosity or thick-disk regime $\alpha < H/R$, and diffusive in the high-viscosity or thin-disk regime $\alpha > H/R$.
The characteristic propagation time $t_{\rm warp}(R) = R/v_{\rm warp}$ is always shorter than the viscous time $t_\nu(R)$. We caution that global simulations of warped disks (Sorathia et al.\ 2013) indicate more complicated dynamics, such as a nonlinear dependence of propagation speed on the strength of the warp.  Equation (\ref{eq:v_warp-warp-velocity}) nevertheless provides a convenient prescription on which to base our discussion.

It is often possible for there to exist an inner region in which precession outpaces warp propagation ($\Omega_{\rm LT}>t_{\rm warp}^{-1}$).  A steady state then exists in which the disk aligns with the BH equator over these radii, but changes orientation at larger radii to match some external plane; this is the Bardeen-Petterson (B-P) configuration (Bardeen \& Petterson 1975).   However, if the inner disk is sufficiently thick then the hole-aligned region need not exist.
Nelson \& Papaloizou (2000) find that the disk zone which aligns with the BH equator
disappears entirely for midplane Mach numbers less than five ($H/R\gtrsim 1/5$) in the inner region.
This criterion, which is confirmed in numerical simulation of thick disks by Fragile \& Anninos (2005) and Fragile et al. (2007),
implies that there is no inner aligned zone when the inner disk is advective ($H/R\simeq1$).

In our theory, advective regions in TDE disks are never any younger than the viscous time at their outer edges (apart from a transient phase of duration $t_{\nu0}$).   Because warps propagate more rapidly than viscous diffusion, we infer that {\em advective TDE disks, and advective zones within TDE disks, always precess as solid bodies}.    The same conclusion holds for any disk in self-similar expansion, even radiative disks, because the self-similar state is marked by a disk age approximately equal to $t_\nu(R_o)$.

The situation is not so clear for radiative zones after the advective-to-radiative transition, however, because $t_{\nu,{\rm gas}}$ can be much longer than the current age.   To handle this case, we assume the disk is broken into an inner region undergoing solid-body precession with an angular frequency $\Omega_d$ and an outer region which does not couple well enough to participate in this motion.  The boundary $R_{\rm sb}$ between these regions is the largest region through which a warp can propagate in a precession time, i.e.,
\beq\label{eq:SolidBodyRegion}
R_{\rm sb} \Omega_d(R_{\rm sb}) = v_{\rm warp}(R_{\rm sb}).
\eeq

To evaluate this criterion, we need an expression for $\Omega_d(R)$: the rate at which the disk within $R$ would precess as a solid body.
For solid body precession, the centrally-concentrated frame dragging torque, whose magnitude is
\beq
\tau_d(R) = 2\pi L M^{1/2} \sin(\theta) \int_{R_i}^R R'^{-3/2} \Sigma(R') \, dR'
\eeq
(where $\theta$ is the inclination angle between disk and hole) acts upon the disk, whose angular momentum
\beq
J_d(R)= 2\pi M^{1/2} \int_{R_i}^R R'^{3/2} \Sigma(R')\,    dR'
\eeq
has its greatest contributions from large radii.  If the disk precesses as a solid body within $R$, it does so at the rate
\beq\label{Omega_precession}
\Omega_d(R) = {\tau_d(R)\over J_d(R)\sin(\theta)} = 2{\mathbf L} {\int_{R_i}^{R} R'^{-3/2} \Sigma(R')\, dR' \over \int_{R_i}^{R} R'^{3/2} \Sigma(R')\, dR'}.
\eeq

We have ignored any torque transmitted to some outer, non-solid-body region.  The correction for this external torque should usually be small, except in cases where $R_{\rm sb}$ divides two very different regions.  It may be significant for an advective disk precessing within a larger radiative disk, however.  We have also ignored any wind torque, but this is justified because the disk and its wind share a common axis.  Finally, we have ignored the torque due to the incorporation of fresh material through fallback.

The precession rate $\Omega_d$ can be obtained in closed form for simple surface density profiles such as truncated power laws
(e.g., Liu \& Melia 2002; Fragile et al. 2007).
For instance, when $a= 0.9$, $n=1/2$, $s=1$, $r_i= 1$ and $r_o= 20$, taking $\Sigma \propto r^{s-n}$ in the range $r_i<r<r_o$, and zero otherwise,  gives a precession period of $0.6\times10^5~ M_6$ s.   Furthermore, so long as $r_i$ is fixed and is $\ll r_o$, and so long as the form of $\Sigma(r) \propto r^{-\zeta}$ stays fixed with $-{1/2}<\zeta<{5/2}$, Equation (\ref{Omega_precession}) implies $\Omega_d \propto R_o^{-(5/2-\zeta)}$: precession rate slows down as the advective disk spreads outward.

This simple result is only an approximation, however, when $\Sigma(r)$ has a more complicated structure punctuated by infall and by a transition in its thermodynamics.  In this case it is much more accurate to evaluate the time evolutions of $\tau_d$ and $J_d$ separately.

The torque $\tau_d$ depends strongly on the disk surface density near its inner edge, and this, in turn, is most sensitive to the current rate of accretion.  So long as steady-state accretion has been achieved in the inner disk, this inner profile is given by
\[\Sigma(R) \simeq {\Mdot_{\rm acc}(R_f) t_\nu(R_f) \over 2\pi R_f^2}  \left(R_f\over R\right)^{\zeta_i}\]
(ignoring the correction factor for a wind lever arm: see Equation (\ref{eq:vR-general})), where $\zeta_i=1/2-s$ if the inner disk is advective, and $\zeta_i = 3/5$ if the inner disk is radiative and gas-pressure dominated.  Then, extending the integral to infinite $R$, and assuming $\zeta_i>-1/2$ so that the torque is indeed concentrated at small radii,
\beq\label{eq:torque-from_Mdot_i}
\tau_d \simeq { L (M/R_i)^{1/2} \sin(\theta)\over \zeta_i-1/2} {\dot M_{\rm acc}(R_f) t_\nu(R_f) \over R_f^{2}} \left(R_f\over R_i\right)^{\zeta_i}
\eeq
for all $R$ much larger than $R_i$.

A key point is that $\dot M_{\rm acc}(R_f)$ is the total accretion rate at $R_f$ from all sources: it contains a contribution from both the outer disk and, in the misaligned case, a fallback stream.   This means that expanding, misaligned disks can either have $\tau_d\propto t^{-\eta}$, if the disk contribution dominates the central accretion, or $\tau_d\propto t^{-5/3}$, if fallback dominates.   Expanding aligned disks have $\tau_d\propto t^{-\eta}$, at least during their self-similar expansion phases.   Radiative disks have constant torques for times less than $t_{\nu,{\rm gas}}$ -- either because their mass accretion rate is constant, or because there has been no time for viscous readjustment.

The disk angular momentum $J_d$, which is essentially conserved in the absence of infall or outflow of matter, changes with time when these effects are  present.  For an advective disk undergoing self-similar expansion, $J_d\propto t^{-\eta_J}$ where $\eta_J$ is given by Equation (\ref{etaJ-eta_relation}) with $n=1/2$.    Combining this information we have, for any advective disk in self-similar expansion, a precession rate law $\Omega_d \propto \tau_d/ J_d \propto t^{-\eta_\Omega}$ with
\beq \label{eq:Omega_d_SelfSim}
\eta_\Omega = \begin{cases}
\eta_J - \frac53 =\frac23 {f_j s \over 1+2s(1-f_j)}-\frac53 & ~~ \mbox{misal'd, $\eta>5/3$},\\
\eta_J-\eta = -{4+2s\over 3}, & ~~ \mbox{otherwise},
\end{cases}
\eeq
where $f_j$ is the wind lever arm discussed in Appendix \ref{Sec:app-ss}.  (It is quite likely that precession could be observable even if the disk is `aligned' so far as its evolution is concerned.)

In TDEs we encountered self-similarly expanding advective disks in both the aligned case (prior to $t_{\rm contr}$) and the misaligned case, and the influence of fallback is somewhat different in the two scenarios.   For an aligned disk, newly-arriving fallback is incorporated at $R_o$ and its angular momentum is shed along with that of the entire disk, as in Equations (\ref{eq:M-ode-aligned}).   The original, spreading disk therefore dominates $J_d$ at late times if $\eta_J<2/3$; otherwise $J_d$ is dominated by recent fallback.    Recall that aligned advective disks can undergo contractions due to the combined influence of winds and infall; this is associated with a precipitous drop in $J_d$ at $t_{\rm contr}$.

For a misaligned disk, newly-arriving matter deposits its angular momentum at $R_f$, where the viscous time is relatively short.   The linearity of our windy disk Equation (\ref{eq:DiskEvlnWithWind}) implies that each new contribution to $J_d$ made at time $t_{\rm fb}$ fades as $(t-t_{\rm fb})^{-\eta_J}$, and this means that fallback will not affect $J_d(t)$ for $\eta_J <1$.   If $\eta_J\geq 1$, the decline of $J_d$ is slowed by the recent addition of material.

Finally, we must consider the advective-to-radiative transition and the phenomena associated with it.   In the aligned case, the outcome depends on whether the advective disk experienced a contraction prior to $t_{\rm tr}$.  If not, then it transitions to a radiative disk once and for all, so that $J_d$ and  $\tau_d$ become fixed.    The disk should then develop a B-P configuration as precession outpaces warp propagation in its inner regions.  If it has contracted, however, then the advective cycles create brief episodes in which the inner disk precesses as a solid body, with $\Omega_d$ declining each time it spreads before transitioning back to the radiative state.

For a misaligned disk, the outer region is advective prior to $t_{\rm tr}$, and the inner region fed by fallback can persist in an advective state until $t_{\rm tr,i}$.  Because the outer disk should decouple from the inner one, this transition marks a sudden drop in $J_d(R_{\rm sb})$ and an associated sudden increase of $\Omega_d$.   If the inner advective region acts as a freely-precessing solid body with definite radius $\sim R_f$, then its precession rate should be roughly constant until it, too, transitions to the radiative state.   On the other hand, if a region of the radiative outer disk is coupled to the region which precesses as a solid body, then $\Omega_d$ may change with time.

To assess this possibility we estimate the outer disk surface density at $t_{\rm tr}$, and extrapolate it inward toward $R_f$ according to the $\Sigma(R)/\Sigma(R_o) = (R/R_o)^{-\zeta}$.  Here $\zeta=1/2-s$ if the advective-to-radiative transition preserved the advective disk structure, but it is quite likely that the transition leaves behind a structure characterized by a larger value of $\zeta$.   We then use the properties of a radiative, gas-pressure dominated region to obtain the two possible profiles of $v_{\rm warp}$ from Equation (\ref{eq:v_warp-warp-velocity}); this exercise shows diffusive propagation holds in radiative regions outside $R_f$ for all realistic value of $\zeta$.   Appealing to condition (\ref{eq:SolidBodyRegion}), the zone of solid-body precession extends outside $R_f$ only for precession periods $2\pi/\Omega_d$ in excess of $2\pi R_f/v_{\rm warp}(R_f)$.  This critical period is of order ten years for $\beta\sim1$ and $-1/2<\zeta<0$, but as short as a month if $\beta\sim10$ and $1/2<\zeta<3/2$.  (The minimum period scales as $\beta^{-(0.87 + 0.42\zeta)}$.)

For precession periods long enough that part of the radiative zone participates in solid-body precession,  we find $2\pi/\Omega_d\propto \tau_d^{(6+4\zeta)/(9-10\zeta)}$: the period lengthens with decreasing torque for $-2/3<\zeta<9/10$.    We note that the limit of this behavior for large $\zeta$ is  (period)$\propto \tau_d^{-2/5}$, which, for $\tau_d\propto \dot M_{\rm fb}\propto t^{-5/3}$, would yield $\eta_\Omega=-2/3$, i.e., (period)$\propto t^{2/3}$.   This is the only limit in which this analysis can be relevant to misaligned disks with advective inner regions and radiative outer zones, because for small values of $\zeta$ the radiative zone outside $R_f$ cannot enter solid-body precession before the entire disk becomes radiative.

%%%%%%%%%%%%%%%%%%%%%%%%%%%%%%%%%%%%%%%%%%%%%%%%%%%%%%%%%%%%%%%%%%%%%%%%%%%%%%%%%%%%

\section{Application to Sw J1644+57}\label{S:SwJ1644}

\begin{figure}
\centerline{
\includegraphics[width=9.4cm, angle=0]{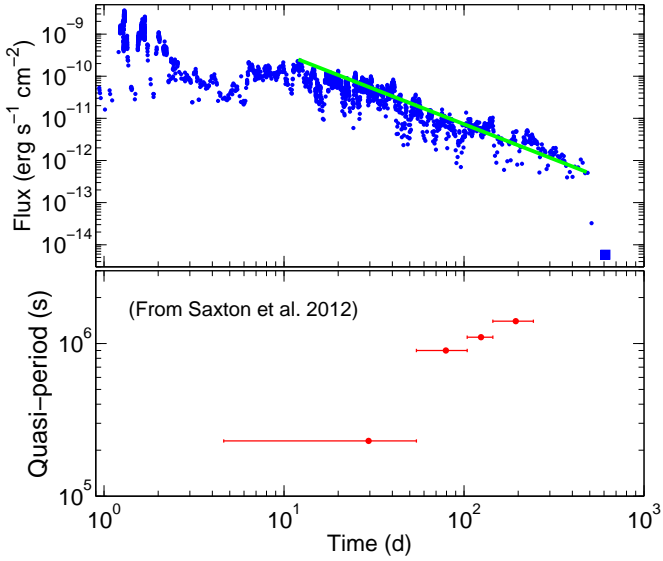}
}
\caption{\textit{Top}: The 0.3 - 10 keV light curve of Sw J1644+57. The Swift XRT data, downloaded from UKSSDC website, are plotted as points. The Chandra observation at $t= 610$ days (Levan \& Tavir 2012; Zauderer et al. 2013) is plotted as a square. The solid green line is a $t^{-5/3}$ power law. 
\textit{Bottom}: The most prominent quasi-periods in four consecutive sections of X-ray light curve found by Saxton et al. (2012) via the Lomb-Scargle periodogram. The horizontal bar of each quasi-period represents the duration of each light curve section.
}  \label{fig:swj1644}
\end{figure}

The recently discovered X-ray transient Sw J1644+57 is a remarkable event that has been identified as a jetted TDE
by several lines of evidence (Levan et al. 2011; Bloom et al. 2011; Burrows et al. 2011; Zauderer et al. 2011). The long-term X-ray light curve of Sw J1644+57 is shown in the top panel of Figure \ref{fig:swj1644}.
It contains multiple flares before $t= 6$ days, then it shows numerous dips thereafter. At a redshift of 0.35, this X-ray transient has an isotropic equivalent luminosity of $10^{47}$ erg s$^{-1}$ during the first 10 days. The super-Eddington luminosity (for a black hole mass $\sim 10^{6-7} M_{\odot}$) together with the sharp variabilities in the light curve suggest that most likely the X-ray photons are directly emitted from a beamed jet, moving at relativistic speed and pointing toward the observer. {The jet is likely to be generated by the Blandford-Znajek mechanism for a black hole with modest to high spin (Lei \& Zhang 2011; Krolik \& Piran 2012).}

Punctuated by dips, the light curve after $t= 13$ days starts a long-term power-law decline consistent with $t^{-5/3}$. This resemblance to the canonical TDE fallback decay power law implies the jet kinetic luminosity might be closely related to the accretion rate at the inner boundary of the disk (assuming the latter also follows the fallback decay power law). During the early time of the light curve ($t< 6$ days), the much more violent behavior there might correspond to the dynamical process of forming the disk and the onset of the jet activity. Models for the early flares have been proposed (e.g., Krolik \& Piran 2011; Wang \& Cheng 2012; Tchekhovskoy et al. 2013).

%%%%%%%%%%%%%%%%%%%%%%%%%%%%%%%%%%%%%%%%%%%%%%%%%%%%%%%%%%%%%%%%%%%%%%%%%%%
%%%%%%%%%%%%%%%%%%%%%%%%%%%%%%%%%%%%%%%%%%%%%%%%%%%%%%%%%%%%%%%%%%%%%%%%%%%

\subsection{Power-law decline and jet shutoff}

The most striking feature of the X-ray light curve is the steep falloff at $t= 500$ d, where the flux dropped by about a factor of 170 on a time scale of $\delta t/t \lesssim 0.2$. Following a speculation by De Colle et al. (2012), Zauderer et al (2013) and Tchekhovskoy et al. (2013) attributed this drop-off to a turn-off of the jet when the mass accretion rate drops below the Eddington rate, with the mechanism unknown.  We agree in general terms with this association, but argue more specifically that it is the transition of the inner disk from an advective to a radiative state which stimulated this drop of emission. The transition is associated with a sharp drop in the disk pressure and scale height, and with a drop in accretion rate by more than a factor of $10^5$ (see Figs. \ref{fig:scandisk} and \ref{fig:with-fallback-aligned})!

Within our model, the time of the transition matches the outer disk transition time $t_{\rm tr}$ if the disk is reasonably aligned with the black hole equator, or at the inner disk transition time $t_{\rm tr,i}$ if it is misaligned.   Both of these are typically earlier than the point at which $\dot M_{\rm fb}$ drops to the Eddington rate $\dot M_{\rm crit}$ (taking a mass-to-radiation efficiency factor of 1, as did Zauderer et al.); in the latter case, this follows from the fact that the fallback radius $R_f$ exceeds the inner disk radius $R_i$.

A self-consistency check can be done for this explanation. First, the start of the power-law decline in light curve at $t= 13/(1+z)$ d should correspond either to the end of circularization $n_{\rm cir} t_f$, or if circularization is rapid ($n_{\rm cir}=1$), to the characteristic fallback time $t_f$ itself. Therefore, 
\beq 	\label{eq:J1644-tf-constraint}
M_6= 0.056~m_*^2 r_*^{-3} n_{\rm cir}^{-2} (t_f/t_*)^{-2}.
\eeq

Second, since we will interpret the dips as a sign of disk precession (see below), we favor a scenario in which the disk and hole are significantly misaligned.  In order for the light curve to follow the $t^{-5/3}$ power law, we infer fallback dominates the central accretion rate, so $\eta>5/3$: this requires that the disk wind parameters be above the solid blue line in Figure (\ref{fig:selfsimdisk}).  This cannot be accomplished if there is no wind from the advective portions of the disk.  If the wind is weak (its mass-loss rate parameter $s<1/4$) then a lever arm is required, but if it is strong ($s\geq1/4$), this is not necessary.

Third, we associate the time of the sharp decline at $500/(1+z)$ days with the advective-to-radiative transition in the inner disk. For the misaligned case using Equation (\ref{eq:t_tr_innerdisk}) for $t_{\rm tr,i}$, we obtain
\beq 	\label{eq:J1644-ttr-constraint}
\beta= 2.7~m_*^{-2/3} (t_f/t_*)^{-2/3}.
\eeq

We then use $t_f/t_* \approx 2 - 3$ found by Guillochon \& Ramirez-Ruiz (2013) for full diruptions (shown in Figure \ref{fig:Tf_Beta_Fit_RR13}), recalling their simulation is appropriate for the Newtonian case ($R_p \gg R_s$) at high mass ratio ($\log_{10}(M/M_*)\sim 6$). Then Equation (\ref{eq:J1644-ttr-constraint}) alone gives $\beta= [1.3, 1.7] m_*^{-2/3}$, meaning that the star is probably fully disrupted, but is still not a deep penetrating event. This conclusion does not explicitly depend on the black hole mass or the stellar type. With this $\beta$, the disruption is Newtonian for $M_6 \lesssim 10$; Sw J1644+57 is well within this mass range (see below).

With $t_f/t_* \approx 2 - 3$, Equation (\ref{eq:J1644-tf-constraint}) alone suggests that, if the star is solar ($m_* \approx r_* \approx 1$), then $M_6 \sim 0.01$ for a rapid circularization ($n_{\rm cir} \sim 1$); a slower circularization ($n_{\rm cir} > 1$) leads to an even lower BH mass. 

This latter finding is in tension with our expectation that the black hole should be at least an order of magnitude more massive. Our BH mass consraint derives from the fact that the required $t_f$ for Sw J1644+57 is unusually short (Equation \ref{eq:J1644-tf-constraint}).  A more massive star alleviates the problem somewhat, especially if the stellar metallicity is low, because these stars are more compact.  For instance, using the zero-age main-sequence mass-radius relation (Tout et al.\ 1996), a star with $m_*=3$ and metallicity a tenth of solar has $m_*/r_*=2$, giving $M_6= 0.07~n_{\rm cir}^{-2}$. For higher stellar masses, the disruption would be only partial ($\beta <1$ from Equation \ref{eq:J1644-ttr-constraint}) and this leads the advective-to-radiative transition to be too early.   More compact phases of stellar evolution, such as Wolf-Rayet stars, are too rare to be plausible, and a white dwarf disruption is inconsistent with our constraints. 

In conclusion, the most likely scenario for Sw J1644+57 consistent with our model involves the full disruption of a star of 1-3 $M_\odot$ by an intermediate-mass [(1-7)$\times 10^4 M_\odot$] black hole. This BH mass is significantly below the upper limit ($M \lesssim 10^7 M_{\odot}$; Burrows et al. 2011; Levan et al. 2011) inferred from the $M$-$L_{\rm bulge}$ relation (e.g., Gultekin et al. 2009). It is slightly smaller than the value ($M \sim 10^5 M_{\odot}$) inferred via other methods (Miller \& G\"{u}ltekin 2011; Reis et al. 2012; Abramowicz \& Liu 2012).

We could also interpret the light curve in our aligned-disk scenario. If the disk has undergone a radial contraction, then its radius at the advective-to-radiative transition is $F_w^{-2} R_f$, and Equation (\ref{eq:t_tr-post-contr}) applies; the solution for $\beta$ is increased by a factor $F_w^{-2}$ relative to what we found above.  This scenario would produce a constant precession period, however, which does not explain the dips (see below).

Our finding of a mild disruption ($1\lesssim \beta\lesssim 2$) is in contrast with those by Cannizzo, Troja \& Lodato (2011) and Gao (2012). With data available only up to $t\approx 100$ d, Cannizzo et al.\ identified $t_f$ to be $< 1$ d, from which they used $t_f\propto \beta^{-3} t_*$ (as opposed to that in Figure \ref{fig:Tf_Beta_Fit_RR13}) 
and assumed $M_6=1$, to obtain  $\beta \approx 10$.  Our $\beta$ is obtained from associating the apparent X-ray shutoff with the disk advective-to-radiative transition, an independent constraint. Given that the $t^{-5/3}$ luminosity decline starts at 13 d, and that $t_{\nu0}/t_f < 1$ for nominal parameters so that a long dynamical delay is unlikely,  we believe our inference $t_f \simeq 13/(1+z)$ d to be robust.

As for the very bright flares before $t=$ 2 d, we consider these most likely due to violent dissipation in the circularization phase of the most-bound debris, e.g,  stream-stream collisions near the pericenter.  We note that Haas et al.\ (2012) observe prompt accretion just after pericenter passage, which could cause an early start-up of the jet activity.
Regardless of what causes the flares, any order-of-magnitude variations in the early light curve should occur on time scales not shorter than the internal dynamical time of the star $t_{\rm sd}= 2\pi [R_*^3/(GM_*)]^{1/2}= 10^4~(r_*^3/m_*)^{1/2}$ s -- comparable, for a solar-type star, to the durations of the very bright flares at $t<$ 2 d. The circular orbital time at the pericenter radius is smaller than $t_{\rm sd}$ by a factor of $\beta^{3/2}$.

\subsection{Dips}

The dipping feature that punctuates the power-law decay of the light curve is intriguing. Time resolved spectral analysis shows no evidence of increased X-ray absorbing column density when the dips occur (Burrows et al. 2011, supplemental information), meaning it is unlikely due to episodic obscuration. Marginal evidence for periodicity is found (Burrows et al. 2011; Saxton et al.\ 2012; Lei et al. 2013). In particular, Saxton et al.\ (2012) searched for periodicity in four consecutive sections of the late X-ray light curve and found signs of quasi-periodicity at multiple periods. The most prominent quasi-periods in each section are 0.23 Ms, 0.9 Ms, 1.1 Ms and 1.4 Ms, respectively, and they are plotted in Fig. \ref{fig:swj1644}. The increase of quasi-periods with time agrees with a visual inspection of the light curve that the time interval between dips are larger at later times.   The quasi-period grows roughly as $t^{2/3}$.

We interpret the dips as being modulation of the jet luminosity by the disk precession. The modulation can be done in two possible scenarios: (1) The jet is affected or deflected by the disk wind, so that its emission is enhanced in the plane which includes the black hole spin axis and the disk normal, similar to the scenario proposed for the ultra-luminous X-ray source SS433 (Begelman, King \& Pringle 2006); as the disk precesses, the jet does do, moving in and out of observer's line of sight. (2) As a misaligned disk precesses, the fallback stream hits the disk outer edge twice in each full precession, and this temporarily reduces the accretion rate near the black hole.

The first of these is similar to the scenario proposed by Lei et al.\ (2013).  Lei et al.\ associate the dips with L-T precession at a single radius, which they associate with the B-P radius.  We have argued in  \S \ref{S:Precession} that during the jet-driving phase there exists a solid-body region of disk precession, and the precession period involves this region's angular momentum as well as the relativistic torque.   For the particular scenario we favor to fit the overall light curve -- that of a significantly misaligned disk created by a moderate plunge ($1\lesssim \beta \lesssim 2$) -- we found that, before the outer disk transitions to radiative, the precession period grows as $t^{\sim 1}$, e.g., for $\eta > 5/3$, $f_j=1$ and $s=1$ (Equation \ref{eq:Omega_d_SelfSim}); after $t_{\rm tr}$, part of an outer, radiative region can precess in step with the inner advective zone, and that this affects periods early enough to be observed provided that the radiative region has a relatively steep density profile, $\zeta\sim1.5$.  Perhaps fortuitously, we found that the period dependence tends to $t^{2/3}$ in this case, as this is the trend of quasi-periods in Sw J1644+57.

%%%%%%%%%%%%%%%%%%%%%%%%%%%%%%%%%%%%%%%%%%%%%%%%%%%%%%%%%%%%%%%%%%%%%%%%%

\section{Summary and Discussion}

In TDE modeling, it has often been assumed that the accretion rate history, hence the emission light curve, tracks the rate at which new debris mass falls back onto the disk, at a rate proportional to $t^{-5/3}$ at late times.  This need not be true, however, when the evolution of the fallback disk due to viscous spreading is considered.  The possibility of a spreading disk is associated with a number of physical effects, such as changes in the black hole accretion rate, disk winds, thermal instabilities, disk-fallback interaction, and precession, all of which complicate the physical picture considerably.  If the observable signals from TDEs can be deciphered to provide constraints on these phenomena, one should gain important knowledge about the physics of transient disks which evolve through a wide range of dimensionless accretion rates.

Our contribution has been to address questions of disk evolution using the simplest self-consistent parameterizations and physical models we can construct.   The elements include a simplified model to capture the dynamics of fallback (\S \ref{S:TDEbasics}); a consideration of the possible thermal states of radiative, radiation-pressure dominated disks and an assessment of the likeliest outcome given the current state of numerical simulations (\S \ref{SS:viscosity}, \S \ref{SS:StateTransitions}); a model for the evolution of a disk without continuing fallback (\S \ref{sec:no-fallback}, which may also be applicable to transient disks in compact-object mergers); an assessment of the influence of fallback in the aligned (\S \ref{sec:Aligned}) and misaligned (\S\ref{subsec:Misaligned}) cases; estimates regarding the dynamics of disk precession (\S \ref{S:Precession}), and application to the source Sw J1644+57 (\S \ref{S:SwJ1644}).  We have relied at several points on a new, self-similar model for the structure and evolution of windy advective disks (Appendix \ref{Sec:app-ss}) and on a Green's function analysis of the response of a spreading disk to the addition of fallback at its inner radii (Appendix \ref{sec:app-green}).

Our models are necessarily approximate, and rely on idealizations.  One of these is our analytical approximation to the thermal and viscous properties of the disk.  Another is the assumption that disks can be neatly divided into aligned and misaligned states, and that these interact quite differently with the stream of fallback material.  A third is our assumption that an inner zone of solid-body precession responds freely to the Lense-Thirring torque and is relatively unaffected by matter orbiting outside its edge.   All of these caveats, which provide avenues for further improvement, render our results somewhat tentative.

Nevertheless, we are encouraged that our  models yield apparently sensible results when applied to Sw J1644+57.  For a black hole mass and a stellar type within the range of what is expected in this source (Levan et al. 2011), the start of the power-law decline of the luminosity and the sudden extinction of the source, as well as the power-law slope connecting these events, are all consistent with our expectations in the case that the star's orbit was misaligned with the black hole spin plane and that its plunge was relatively deep without being relativistic.   Perhaps fortuitously, the same scenario can produce the trend in precession period seen in this source (period $\propto t^{2/3}$) under reasonable assumptions about the structure of disk material left behind by an early epoch of viscous evolution.  These possibilities merit more detailed scrutiny than we provide here.   It is important to state, however, that for a fixed black hole mass and specific angular momentum of the returning material, we know of no physical mechanism other than an evolving zone of locked precession which could give rise to a precession period which increases in time.

After this paper was submitted, Kawashima et al.\ (2013) presented simulations of the global limit cycle behavior of a radiation pressure dominated disk, aiming to explain the sudden jet shutoff of Sw J1644+57.  In contrast to our models, these authors assume the disk is fed at its outer boundary $R_o \approx 100 R_S$ with a constant mass supply rate (in units of $\dot{M}_{\rm crit}$) $\approx R_o/R_S$.  This high and constant mass feeding rate strongly overestimates the chances of a jet revival (see our Equation \ref{eq:t_tr_i-compared-to-t_1} and related discussion in Section \ref{sec:Aligned}).  Even in cases where a revival occurs,  assuming a high, constant accretion rate underestimates the delay time at which it occurs. Moreover, assuming such a large feeding radius $\gg R_f$ leads to an overestimate of the advective phase duration of any limit cycle. Note that the speculation of jet revival in Tchekhovskoy et al. (2013) is not due to the limit cycle behavior, and they assumed that the late mass accretion rate tracks the fallback rate, which we have shown should not be the case in TDEs.

Our analysis has two fundamental points. The first regards the importance of advective disk winds for the observational properties of tidal disruption events.  In addition to strongly modulating the emission (Strubbe \& Quataert 2009, 2011), winds' dynamical influence makes it possible for the BH accretion rate to follow the $t^{-5/3}$ time dependence of stellar fallback.  In the case of an aligned disk, we have seen that this can occur because the combined influence of winds and fallback can lead to a contraction of the disk radius, leaving behind a compact steady-state structure which promptly processes what falls upon it.  For misaligned disks, we have argued that it occurs because wind suppresses the central accretion rate from a spreading outer disk, which would otherwise dominate the central accretion rate before the outer disk transitions to a radiative state.

The other fundamental point regards the instability of radiative, radiation pressure dominated disks. A distinctive feature of the Shakura \& Sunyaev prescription for local dissipation, this instability is still a major unresolved issue in the theory of accretion disks.  Our interpretation of the jet shutoff in Sw J1644+57 requires the instability to operate.
However, many black hole X-ray binary (BHXBs) and active galactic nucleus (AGN) systems accrete at or above Eddington rates so that the inner region of their disks should have entered in the unstable regime (ii), but only a few show strong limit-cycle like flux variations; see Done et al.\ (2007) for a review. The best of these few cases is GRS 1915+105 (Taam, Chen \& Swank 1997; Fender \& Belloni 2004). A mixed parameterization, in which $\nu \propto \alpha P_{\rm gas}^{\delta} P^{1-\delta}$ has been proposed to resolve this (Honma et al. 1991; Merloni \& Nayakshin 2006; Czerny et al. 2009) and the instability exists for $\delta < 4/7$ (Kato et al. 1998).   A complication is that AGN and BHXB disks extend to much larger radii, relative to $R_S$, than do TDE disks; this affects the nature of any global thermal cycles, and adds additional physics such as the hydrogen ionization instability (e.g., Janiuk \& Czerny 2011).

Within the range $0<\delta<4/7$ an increase of $\delta$ causes the advective regime to shift toward higher $\Sigma$ (Figure \ref{fig:scandisk}), reducing in amplitude the change of $\dot{m}_{\rm acc}$ during the transition from regime (i) to regime (iii), or vice versa.  The end result is that the transitions which lead to limit cycle behavior become less significant, disappearing entirely for $\delta>4/7$.   We expect that the jet shutoff in Sw J1644+57 can be accommodated by these models, at least up to a maximum value of $\delta$; indeed our predictions for observables will hardly change because the advection-to-radiative transition criterion $(\dot{m}_{\rm acc})_{\rm i-ii}= r$ remains the same, and because the accretion rate follows the fallback rate as long as the wind loss is strong ($\eta > 5/3$ for the misaligned case, $\eta_{\dot{M}_d} > 5/3$ for the aligned case).   While {mixed prescriptions} with finite $\delta$ may fit observed transition {of Sw J1644+57}, the physics of {radiation pressure} dominated disks are likely to be much richer than can be represented by any analytical viscosity law.

There are several immediate avenues for further investigation of evolving TDE disks.  One is to model the emission of relatively long-wavelength thermal radiation from the spreading advective disk and its wind, as well as from the subsequent radiative disk, or the variation of linear polarization over its precession, to provide observational tests and diagnostics of its presence. Sw J1644+57 is luckily a special case whose luminosity is dominated by X-rays emitted from a jet pointing toward us; in this case the jet power, hence the X-ray luminosity, traces the accretion rate at the inner boundary of the disk. Almost all previous calculations of TDE disk light curves have assumed accretion onto a disk of fixed radius tapering off as $t^{-5/3}$, but we have found that one or both of these assumptions can be incorrect in either the early advective state, or the later radiative state. Another is to extend Greens-function solutions for the disk evolution to the case where the disk blows a wind, and in which the wind angular momentum is enhanced by a magnetic lever arm; the linearity of our Equation (\ref{eq:DiskEvlnWithWind}) shows that this is possible.  A third would be to consider disk wind models, to determine whether the wind-induced instability we highlight in Appendix \ref{Sec:app-ss} is avoided by astrophysical disks, or could instead be a source of intermittency in accretion systems.

\section*{}
This work is supported by NSERC through a Discovery Grant.  The authors are indebted to Linda Strubbe, Shane Davis, Brian Metzger, Yi Feng, Wei-Hua Lei, Nick Stone, James Guillochon, Julian Krolik and Zhuo Li for insightful and helpful discussions. This work made use of data supplied by the UK Swift Science Data Centre at the University of Leicester.

%%%%%%%%%%%%%%%%%%%%Begin the Reference%%%%%%%%%%%%%%%%%%%%%%%%%%

%%%%%%%%%%%%%%%%%%%%End the Reference%%%%%%%%%%%%%%%%%%%%%%%%%%%

%%%%%%%%%%%%%%%%%%%%%%%%%%%%%%%Begin the Appendix%%%%%%%%%%%%%%%%%%%%%%%%%%%%%%%%%%%%%%%%%%%%%%%%%%
\appendix

%%%%%%%%%%%%%%%%%%%%%%%%%%%%%%%Begin Appendix B%%%%%%%%%%%%%%%%%%%%%%%%%%%%%
\section{A. Self-simlar evolution of windy, spreading disks without fallback}\label{Sec:app-ss}

To understand the time evolution of an advective disk, we wish to consider the effect of a wind on the evolution of a disk which spreads well beyond the fallback radius.   Because of the rapid decline in the rate of fallback, we consider only the remnant disk from an early period of rapid accretion; however we must then check that the late arrival of matter does not spoil our solution.

Disk matter has mass per unit radius $dM_d/dR = 2\pi R \Sigma$, where $\Sigma(R)=\int_{-\infty}^\infty \rho(R,z)\,dz$ is the total column density, and specific angular momentum $j(R)=(GMR)^{1/2}$ where $M$ is the central mass.  Wind removes mass at a rate $\dot\Sigma_w$ per unit area and removes angular momentum at a rate $f_j j \dot\Sigma_w$ per unit area, where $f_j>1$ if there is any magnetic lever arm.  The mass loss rate per unit radius is $d\dot M_w/dR = 2\pi R \dot\Sigma_w$.  In the presence of a viscosity $\nu$, angular momentum is also redistributed within a Keplerian disk by the viscous torque $g = 3\pi j \nu \Sigma$.  Subtracting the mass conservation equation
\begin{equation}\label{eq:Mass-conservation}
 \ddt \dMdr + \ddr v_R \dMdr + \dMdotwdr = 0
 \end{equation}
 from the angular momentum conservation equation
\begin{equation}\label{eq:AngMom-conservation}
\frac1j \ddt j \dMdr + \frac1j \ddr \left( j v_R \dMdr + g\right) + f_j \dMdotwdr =0
 \end{equation}
 one finds that the radial velocity satisfies
 \begin{equation}\label{eq:vR-general}
 v_R = -{3\over \Sigma} R^{-1/2} \ddr R^{1/2} \nu \Sigma- 2(f_j-1) R{\Sigmadotw\over \Sigma}.
\end{equation}
The viscous accretion rate is $\dot M=-2\pi R\Sigma v_R$.  For steady state regions with no wind or windy regions with no lever arm ($f_j=1$), this gives the familiar expression $\dot M = 3\pi \nu \Sigma$.

For non-radiative disks, a standard form for the wind is one which imposes  $\Mdot(R)\propto R^s$  for regions of steady accretion.   Mass conservation then requires $\Sigmadotw = s\dot M/(2\pi R^2)$ in those regions.   There are at least two expressions for $\Sigmadotw$ which take this limiting form, including $\Sigmadotw = -s \Sigma v_R/R$ and
\begin{equation}\label{eq:Sigmadotw}
\Sigmadotw = sK \Sigma \nu/R^2,
\end{equation}
for some constant $K$ to be determined.  The first of these is not physically motivated and can be negative in the outer disk when $v_R$ is positive.   We adopt Equation (\ref{eq:Sigmadotw}) instead; this  corresponds to
\begin{equation}	\label{eq:wind-energy}
\frac12 R^2 \Omega^2 \Sigmadotw = \frac {sK}9 Q^+,
\end{equation}
so that an unmagnetized wind carries a fraction $(sK/9) v_w^2/(\Omega R)^2$ of the local viscous dissipation if $v_w$ is the terminal velocity of wind originating at $R$.

The value of $K$ can be determined by reference to a steady-state zone of windy accretion.  Setting $sK\Sigma \nu/R^2 = s \dot M/(2\pi R^2)$ and using $\dot M = -2\pi R \Sigma v_R$, we find
\begin{equation} \label{eq:K-for-wind}
K = \frac32 {1+2s\over 1-2s(f_j-1)}
\end{equation}
For a given $\Sigma(R)$, a finite lever arm enhances the steady-state inflow speed by the factor $1/[1-2s(f_j-1)]$, relative to the case in which $f_j=1$. Note that $K$ and the steady-state value of $v_R$ both diverge for $f_j \rightarrow 1+1/(2s)$, and take the wrong sign for all larger values of $f_j$.  This corresponds to an instability in which wind torque stimulates inflow, which induces more wind, and so on.  It is quite possible that real disks can exist in a state of wind-induced instability; alternately, the physics of wind emission may avoid such a state.   For now,
 we restrict our attention to the case of smooth flows, i.e., those with $f_j<1+1/(2s)$.

Using our formula for $v_R$ in the mass conservation equation,
\begin{equation}\label{eq:DiskEvlnWithWind}
\ddt \Sigma =  \frac{3}{R} \ddr R^{1/2} \ddr R^{1/2} \nu \Sigma + {2s(f_j-1) K \over R} \ddr  \nu \Sigma - {sK\over R^2}\nu \Sigma.
\end{equation}

In the case where $\nu\propto R^n \Sigma^q$ a thermally and viscously-stable disk (one with $q>-1$) will tend toward self-similar state in which $\Sigma(R,t)$ can be reduced to powers of $R$ and $t$ times a function of the self-similar coordinate $\xi\equiv R^2/(\nu t)$.    Because we are interested in windy, radiation pressure-dominated, advective disks, we assume $q=0$ in what follows, and then specify $n=1/2$ when tabulating our results.

  We are interested in solutions which extend to the origin with no torque, so that $\Mdot\propto \nu \Sigma \propto R^s$ at small radii, and which have no external source or sink of mass (apart from their own winds) at large radii.   We therefore take $\Sigma = \mbox{const.}\,R^s F(\xi)/(\nu t^{\eta})$, where $\eta$ is an exponent to be determined by our constraints, and where $F$ and its derivatives are finite for $\xi\rightarrow 0$.  With this ansatz, Equation (\ref{eq:DiskEvlnWithWind}) becomes an ordinary differential equation for $F(\xi)$:
\begin{eqnarray} \label{eq:SelfSimilarODE}
A\xi F''(\xi) + (B+\xi)F'(\xi) + \eta F(\xi) &=&0,
\end{eqnarray}
where
\[ A= 3(2-n)^2, ~~ B= (2-n)[3(2+s-n)+K]. \]
Only one solution extends from small to large $\xi$ in the manner of an isolated, spreading disk:
\begin{equation} \label{eq:SelfSimilarEta}
\eta = \frac{B}{A} = 1 + {s\over 2-n} + {1+2s \over (4-2n)[1-2s(f_j-1)] },
\end{equation}
which implies
\begin{equation} \label{eq:SelfSimSolution}
F(\xi) = {\rm const.}\,e^{-\xi/A}.
\end{equation}
For any smaller value of  $\eta$, $F(\xi)$ becomes asymptotically constant at large $\xi$: the disk is infinite, implying a source of mass at large distances.  For any larger value, $F(\xi)$ goes to zero at finite $\xi$ but the rate of mass outflow does not: mass is actively removed at the outer boundary.

We note that Kumar et al (2008) use angular momentum conservation in an approximate model of windy, advective disks with no lever arm to arrive at $\eta = 4(1+s)/3$, and this agrees precisely with Equation (\ref{eq:SelfSimilarEta}) in that limit ($n=1/2, f_j=1$).

Within this solution, the total disk mass $M_d$ and its rate of change $\dot M_d$ vary according to $\dot M_d \propto M_d/t \propto t^{-\eta_{\dot M_d}}$ with
\begin{equation}\label{eta-Mdot}
\eta_{\dot M_d} = \eta - {s\over 2-n} = 1 + {1+2s \over (4-2n)[1-2s(f_j-1)] };
\end{equation}
the disk angular momentum varies as $t^{-\eta_J}$ with \begin{equation} \label{etaJ-eta_relation}
\eta_J = \eta - {5 + 2(s-n) \over 4-2n} = {f_j s \over (2-n)[1-2s(f_j-1) ]}.
\end{equation}
Note that all of these indices diverge at the boundary of the wind-induced instability.

\begin{figure}
\centerline{
\includegraphics[width=9cm, angle=0]{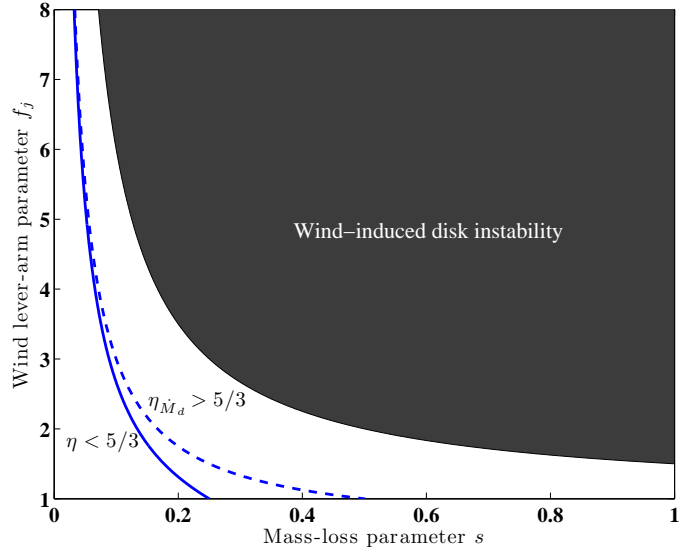}
}
\caption{Parameter space self-similar spreading disks with winds, evaluated for the advective, radiation-pressure dominated case and a Shakura \& Sunyaev viscosity law ($q=0, n=1/2$).  The region of large wind lever arm $f_j$ is excluded due to a wind-induced instability.  The criteria for $\dot M_d \propto t^{-5/3}$ and for $\dot M_{\rm acc}(R_i)\propto t^{-5/3}$, which correspond to $\eta_{\dot M_d} = 5/3$ and $\eta=5/3$, respectively, are shown.}\label{fig:selfsimdisk}
\end{figure}

At several points we are interested in the criterion $\eta<5/3$, for which the central inflow of a spreading disk can come to dominate over the central accretion caused by a fallback stream impinging on the inner disk.  This requires $f_j< (1-4s^2) /[4s(1-s)]$.
In \S \ref{sec:Aligned} we are interested in the criterion $\eta_{\dot M_d} < 5/3$, for which the total disk mass loss rate declines more slowly than the rate of fallback so that self-similar expansion is possible in the presence of accretion at the outer edge. For $n=1/2$ this is true when $f_j < 2/(5s)+4/5$.     We depict this criterion, as well as the boundary of the wind-induced instability discussed above, in Figure \ref{fig:selfsimdisk}.

We have not extended this self-similar analysis to viscosity laws, such as Sakimoto \& Coroniti's, in which $\nu$ depends on $\Sigma$ and $q\neq0$.  Pringle (1991) has shown that when winds are absent, the solutions in this case are, like our Equation (\ref{eq:SelfSimSolution}), very simple functions of $\xi$.  We suspect that windy, self-similar disks with $q\neq0$ are equally simple.

%%%%%%%%%%%%%%%%%%%%%%%%%%%%%%%%Appendix A%%%%%%%%%%%%%%%%%%%%%%%%%%%%%%
\section{B. Green's function solution to the viscous spreading disk with fallback and without wind}

\label{sec:app-green}

Here we show the Green's function solution to the viscous evolution of a disk with misaligned fallback, for $\nu \propto R^n$ which is relevant to the advective state of the disk.   We consider the case in which fallback mass enters the disk with Keplerian angular momentum at a rate per unit area $S(R,t)$, and we consider the case where winds are absent, so $\Sigma$ obeys the viscous diffusion equation
\beq
\ddt \Sigma =  \frac{3}{R} \ddr R^{1/2} \ddr R^{1/2} \nu \Sigma + S(R,t).
\eeq
The solution (see Tanaka 2011 and Metzger et al. 2012 for the derivation) is:
\beq		\label{eq:green-sol}
\Sigma(x, t)= \int_0^{\infty} \Sigma(x', t=0) G(x, x', t) dx' + \int_0^{\infty} dx' \int_0^{t} S(x', t') G(x, x', t-t') dt',
\eeq
with the Green's function
\beq		\label{eq:green}
G(x, x', t)= \frac{(2-n)}{2}\frac{x'^{5/4}}{x^{1/4+n} \tau(t)} I_l\left[\frac{(xx')^{1-n/2}}{\tau(t)}\right] \exp\left[-\frac{x^{2-n} + x'^{2-n}}{2\tau(t)}\right],
\eeq
where $x= R/R_f$, $\tau(t)= (2-n)^2 t/t_{\nu,0}$ is the normalized time by $t_{\nu,0}= 2R_f^2/[3\nu(R_f)]$ the viscous time scale at $R_f$, and $I_l(z)$ is the modified Bessel function of the first kind with the order $l= 1/[2(2-n)]$.

\subsection{Without Fallback}\label{subsec:NoFallback}

For demonstrative purpose, let us firstly examine the simplest case in which there is no fallback ($S=0$) so only the first integral in Equation (\ref{eq:green-sol}) remains, and the disk started as a ring of mass at $R_f$: $\Sigma(x,t=0)= \Sigma_0 \delta(x-1) x$. The solution is the Green's function $G(x, 1, t)$ itself:
\beq		\label{eq:del-green-sol}
\Sigma(x,t)= \Sigma_0 \frac{(2-n)}{2} x^{-1/4-n} I_l\left[\frac{x^{1-n/2}}{\tau}\right] \exp\left[-\frac{x^{2-n} + 1}{2\tau}\right].
\eeq

We may estimate the asymptotic behavior of $\Sigma(x,t)$ by noting the asymptotic form of $I_l(z)$:
\beq		\label{eq:besseli}
I_l(z) \simeq \begin{cases} \frac{(z/2)^l}{\Gamma(l+1)}, & \mbox{for} ~~ z \lesssim 1,\\
							\frac{\exp(z)}{\sqrt{2\pi z}}, &	\mbox{for} ~~ z \gtrsim 1.
							\end{cases}
\eeq
Therefore, the exponential drop-off terms for large $x$ in Equations (\ref{eq:del-green-sol} - \ref{eq:besseli}) determine the outer edge of the disk:
\beq
x_{\rm out}(t)= [1+ 2\tau(t)]^{1/(2-n)}.
\eeq
On the small $x$ limit, one finds $\Sigma(x \lesssim x_{\rm out}, t > t_{\nu,0}) \propto x^{-n} t^{-l-1}$. One can also find the accretion rate by
\beq		\label{eq:dotm}
\dot{M}(x,t)= -2\pi R v_r= 6\pi x^{1/2} \frac{\partial}{\partial x} (\nu \Sigma x^{1/2}),
\eeq
which gives $\dot{M}(x \ll x_{\rm out}, t > t_{\nu,0}) \propto t^{-l-1}$, i.e., it does not depend on $x$. However, as we have shown in Appendix \ref{Sec:app-ss}, this property changes when there is wind.

%%%%%%%%%%%%%%%%%%%%%%%%%%%%%%%%%%%%%%%%%%%%%%%%%%%%%%%%%%%%%%%%%%%%%%%%%%%%%%%
\begin{figure}
\centerline{
\includegraphics[width=10cm, angle=0]{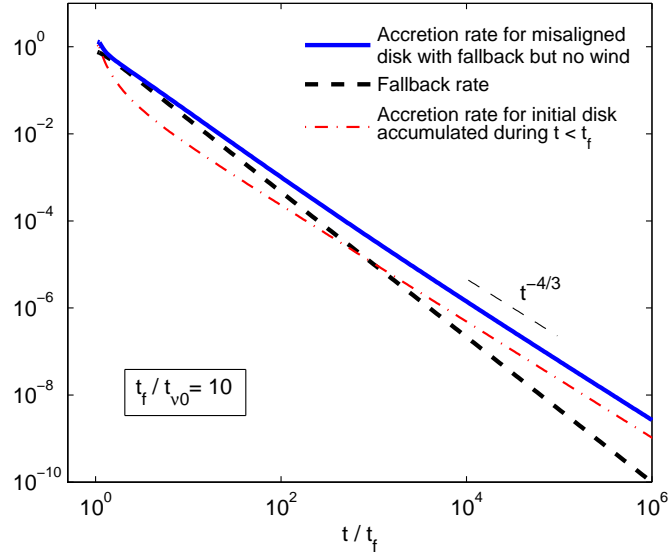}
}
\caption{Numerically calculated accretion rate at $r_f$ for a misaligned disk with fallback but without wind, using Green's function solution Eqs. (\ref{eq:del-green-sol}), (\ref{eq:dotm}) and (\ref{eq:fb-green-sol}). $n= 1/2$ is chosen, appropriate for the early advective accretion regime. The fallback rate is described by Equation (\ref{eq:dotmfb}) but only the declining part is plotted.  The accretion rate approaches the asymptotic $t^{-4/3}$ power law, shallower than the declining fallback rate power law. Also shown is the accretion rate from the initial disk accumulated from early fallback ($t<t_f$), which contributes only partially to the shallower decline of the overall accretion rate. Note that the transition of accretion state to the gas-pressure dominated regime is ignored here, but it is included in \S~\ref{subsec:Misaligned}.} \label{fig:green}
\end{figure}
%%%%%%%%%%%%%%%%%%%%%%%%%%%%%%%%%%%%%%%%%%%%%%%%%%%%%%%%%%%%%%%%%%%%%%%%%%%

\subsection{With Fallback} \label{subsec:Fallback}

Now let us consider the impact of fallback. For simplicity we assume $\Sigma(x,t=0)= 0$, so the first integral in Equation (\ref{eq:green-sol}) disappears. We also assume the fallback material intercepts the disk within a small radial interval, so that
\beq
S(x,t)= \frac{\dot{M}_{\rm fb}(t)}{2\pi R_f^2} \delta(x-1).
\eeq
Then,
\beq		\label{eq:fb-green-sol}
\Sigma(x,t)= \int_0^{t} S(t') G(x, 1, t-t') dt',
\eeq
where $S(t')= \dot{M}_{\rm fb}(t')/(2\pi R_f^2)$.

As shown in Metzger et al. (2012), for the case when $S(t')$ varies slowly one can get a simple solution for $\Sigma(x,t)$ for late times $\tau(t) \gg x^{1-n/2}$. In this case, $S(t')$ can be considered almost constant and be taken out of the integral. Thus,
\bea
\Sigma(x,t) & = & S(t) \int_0^t G(x, 1, t') dt' \nonumber \\
			& = & \frac{S(t)}{2(2-n)}\frac{t_{\nu,0}}{x^{1/4+n}} \int_{\frac{x^{1-n/2}}{\tau(t)}}^{\infty} I_l(z) \exp\left[-\frac{(x^{1-n/2}+x^{n/2-1})}{2}z\right] \frac{dz}{z}. 	
\eea
For late times $\tau(t) \gg x^{1-n/2}$, the integral lower limit can be approximated as $0$. Then utilizing the formula
\beq
\int_0^{\infty} I_{\nu}(u) \exp(-\lambda u) = \frac{1}{\nu} (\lambda+\sqrt{\lambda^2-1})^{-\nu},
\eeq
one obtains
\beq \label{eq:Sigma_Profile_From_GreensFn}
\Sigma(x,t) \approx S(t) t_{\nu,0} \times \begin{cases} x^{-n}, & \mbox{for} ~~ x < 1,\\
													  x^{-n-1/2}, & \mbox{for} ~~ 1< x < x_{\rm out}.
					\end{cases}
\eeq
For the accretion rate, one finds through Equation (\ref{eq:dotm})
\beq
\dot{M}(x,t) \approx 2\pi R_f^2 \times \begin{cases} S(t), & \mbox{for} ~~ x < 1,\\
													 0, & \mbox{for} ~~ 1< x< x_{\rm out}.
					\end{cases}
\eeq
This means at late times and for a slowly varying fallback rate, the central black hole gains mass at the same rate it is supplied at $R_f$, as if there were no viscous outflow.  In fact, there is an outflow associated with this inflow of material; but at late times, the radial outflow of matter from $R_f$ is balanced by the return of matter which previously diffused outward.

The two power law profiles we find in Equation (\ref{eq:Sigma_Profile_From_GreensFn}) are in steady state ($\dot\Sigma=0$) for different reasons: steady-state inflow in the first case, and zero net flow in the second.   Although we do not provide complete solutions for the windy case, it is simple to generalize this finding to the case of a windy disk.  Setting $\dot\Sigma=0$ in Equation (\ref{eq:Sigmadotw}), we find that the newly-added material has $\Sigma\propto R^{s-n}$ for $R<R_f$, steepening to $\Sigma\propto R^{-K/3-n}$ for $R>R_f$, where $K$ is defined in Equation (\ref{eq:K-for-wind}).

For an evolving $S(t)$, one has to numerically calculate the integral in Equation (\ref{eq:fb-green-sol}) in order to get the solution for $\Sigma(x,t)$.  Figure \ref{fig:green} shows the numerical results for the fallback history given by Equation (\ref{eq:dotmfb}). It shows that the accretion rate always approaches the asymptotic $t^{-l-1}$ ($t^{-4/3}$ for $n=1/2$) power law unless the fallback decays shallower than that.

%%%%%%%%%%%%%%%%%%%%%%%%%%%%%%%%End the Appendix%%%%%%%%%%%%%%%%%%%%%%%%%%%%%%%%%%%%%%%%%%%%%%%%%


\begin{references}

\reference{} Abramowicz M. A., Czerny B., Lasota J. P., Szuszkiewicz E., 1988, ApJ, 332, 646

\reference{} Abramowicz M. A., Liu F. K., 2012, A\&A, 548, A3

\reference{} Ayal S., Livio M., Piran T., 2000, ApJ, 545, 772

\reference{} Bardeen J. M., Petterson J. A., 1975, ApJ, 195, L65
% (Lense-Thirring effect causes differential precession in disk (their Fig. 1). Analyze the effects of viscous torques on the configuration of differential precession -- they tend to align the inner disk ($< \sim 10^2 M$) with hole equatorial plane. An isotropic Shakura \& Sunyaev viscosity is considered.)

\reference{} Barres de Almeida U., De Angelis A., 2011, arxiv:1104.2528

\reference{} Begelman M. C., King A. R., Pringle J. E., 2006, MNRAS, 370, 399
% (tilted disk wind/outflow deflects the jet causing the jet precession in microquasar SS433)

\reference{} Begelman M. C., 2012, MNRAS, 420, 2912
% (Anlytical model for wind loss in radiatively inefficient accretion flows)

\reference{} Berger, E.; Zauderer, A.; Pooley, G. G.; Soderberg, A. M.; Sari, R.; Brunthaler, A.; Bietenholz, M. F., 2012, ApJ, 748, 36

\reference{} Blandford R. D., Begelman M. C., 1999, MNRAS, 303, L1

\reference{} Bloom J. S. et al., 2011, Science, 333, 203

\reference{} Burrows D. N., et al., 2011, Nature, 476, 421

\reference{} Cannizzo J. K., Lee H. M., Goodman J., 1990, ApJ, 351, 38
% (Returning tidal debris streams intersect and shock between each other as a result of relativistic precession of their eccentric orbits. This circularizes the streams, on a time scale of the same order as the returning time of the mose bound material. They further calculate the time evolution of the $\alpha$-disk, but assume the disk is thin)

\reference{} Cannizzo J. K., Gehrels N., 2009, ApJ, 700, 1047
% (Time-dependent evolution of a debris disk in different accretion rate regimes, especially in GRB context)

\reference{} Cannizzo J. K., Troja E., Lodato G.,  2011, ApJ, 742, 32

\reference{} Cenko, S. B.; Bloom, J. S.; Kulkarni, S. R.; et al., 2012a, MNRAS, 420, 2684

\reference{} Cenko, S. B.; Krimm, H. A.; Horesh, A.; et al.,  2012b, ApJ, 753, 77
% (Sw-J2058 -- a 2nd possible jetted TDE flare. $L_{X, iso} \approx 3\times10^{47}$ erg s$^{-1}$, radio $L_{iso} \sim 10^{42}$ erg s$^-1$, optically faint not due to huge dust extinction)

\reference{} Chevalier, R.~A.\ 1989, ApJ, 346, 847

\reference{}Coughlin, E.~R. \& Begelman, M.~C., 2013, arXiv:1312.5314

\reference{} Czerny, B.; Siemiginowska, A.; Janiuk, A.; Nikiel-Wroczy\'{n}ski, B.; Stawarz, \L., 2009, ApJ, 698, 840

\reference{} Darwin, C.\ 1959, Royal Society of London Proceedings Series A, 249, 180 

\reference{} Davis, S.~W., Stone, J.~M., \& Jiang, Y.-F.\ 2012, ApJS, 199, 9

\reference{} De Colle F., Guillochon J., Naiman J., Ramirez-Ruiz E., 2012, ApJ, 760, 103

\reference{} Done C., Gierli\'{n} ski M., Kubota A., 2007, Astron. Astrophys. Rev., 15, 1

\reference{} Esquej P., Saxton R. D., Komossa S. et al., 2008, A\&A, 489, 543

\reference{} Evans C. R., \& Kochanek C. S., 1989, ApJ, 346, L13
% (Calculates specific energy distribution in disrupted star, which determines mass fallback rate; has good discussion on circularization of returning debris streams)

\reference{} Fender R., Belloni T., 2004, ARA\&A 42, 317

\reference{} Fragile P. C.; Anninos, P., 2005, ApJ, 623, 347
% (3D GR simulation of thick, tilted, invisid disk torus. Inner disk warps freeze in out to $\sim 7 R_G$ at which the local $t_{LT} \approx t_{cs}$ - the azimuthal sound crossing time, and outside which $t_{LT} > t_{cs}$ and the disk precesses as a solid body. Question: if no viscosity, how is accretion rate possible?)

\reference{} Fragile P. C., Blaes O. M., Anninos P., Salmonson J. D., 2007, ApJ, 668, 417
% (Extension of Fragile \& Anninos 2005 taking into account of MPI as agent of transporting angular momentum. Alignment of disk toward BH equatorial plane does not occur. Little warping is observed. Instead, the unwarped disk precesses uniformly)

\reference{} Gao W.-H., 2012, ApJ, 761, 113

\reference{} Gezari, S.; Basa, S.; Martin, D. C. et al., 2008, ApJ, 676, 944

\reference{} Gezari, S.; Heckman, T.; Cenko, S. B.; et al., 2009, ApJ, 698, 1367

\reference{} Gezari, S.; Chornock, R.; Rest, A.; et al., 2012, Nature, 485, 217

\reference{} Giannios, D., \& Metzger, B. D. 2011, MNRAS, 416, 2102

\reference{} Guillochon J., Ramirez-Ruiz E., Rosswog S., Kasen D., 2009, ApJ, 705, 844
% (Simulation studies vertical compression of the star and the shock breakout signal during the pericenter passage in TDE)

\reference{} Guillochon, J., \& Ramirez-Ruiz, E.\ 2013, \apj, 767, 25

\reference{} Guillochon, J., Manukian, H., \& Ramirez-Ruiz, E.,\ 2013, arXiv:1304.6397

\reference{} Haas R., Shcherbakov R. V., Bode T., Laguna P., 2012, ApJ, 749, 117

\reference{} Hawley, J.~F., Balbus, S.~A., \& Stone, J.~M.\ 2001, \apjl, 554, L49

\reference{} Hayasaki, K.; Stone, N.; Loeb, A., 2012, arXiv:1210.1333

\reference{} Hills, J. G., 1975, Nature, 254, 295

\reference{} Hirose, S., Krolik, J.~H., \& Blaes, O.\ 2009, \apj, 691, 16

\reference{} Honma F., Kato S., Matsumoto R., 1991, PASJ 43, 147

\reference{} Janiuk, A.; Czerny, B., 2011, MNRAS, 414, 2186

\reference{} Jiang, Y.-F., Stone, J.~M., \& Davis, S.~W.\ 2012, ApJS, 199, 14

\reference{} Jiang, Y.-F., Stone, J.~M., \& Davis, S.~W.\ 2013, ApJ, submitted

\reference{} Kato S., Fukue J., Mineshige S., 1998, Black-hole Accretion Disks, Kyoto University Press

\reference{} Kawashima T., Ohsuga K., Usui R., Kawai N., Negoro H., Matsumoto R., 2013, arXiv: 1305.4943

\reference{} Kesden M., 2012, Phys. Rev. D, 86, 064026 

\reference{} Kobayashi S., Laguna P., Phinney E. S., M\'{e}sz\'{a}ros P., 2004, ApJ, 615, 855

\reference{} Kochanek C. S., 1994, ApJ, 422, 508

\reference{} Kohri K., Narayan R., Piran T., 2005, ApJ, 629, 341
% (Neutrino accretion flow at center of supernovae. Accretion is advection-dominated. Develope substantial wind outflow, may reviving stalled supernova shock)

\reference{} Krolik J. H. \& Piran T.,  2011, ApJ, 743, 134

\reference{} Krolik J. H. \& Piran T., 2012, ApJ, 749, 92
% (Has extensive discussion of TDE accretion rate evolution. Differentiate the non-thermal jet luminosity from the thermal disk luminosity and make time-dependent comparison of the two. Application to recent Swift TDE transient Sw-J2058)

\reference{} Kumar P., Narayan R., Johnson J. L., 2008, MNRAS, 388, 1729

\reference{} Lei W.-H., \& Zhang B., 2011, ApJL, 740, L27

\reference{} Lei W.-H., Zhang B., Gao H., 2013, ApJ, 762, 98
% (Analyzed Sw J1644 X-ray LC between 2 days and $2.6\times10^6$ s, found a 0.23 Ms quasi-period; interpreted it as the Lense-Thriring period at the Bardeen-Petterson transition radius)

\reference{} Li L.-X., Narayan R., Menou K., 2002, ApJ, 576, 753

\reference{} Levan A. J., et al., 2011, Science, 333, 199

\reference{} Levan A. J., Tanvir N., 2012, The Astronomer's Telegram, 4610

\reference{} Lightman A. P., 1974, ApJ, 194, 429

\reference{} Lodato G., King A. R., Pringle J. E., 2009, MNRAS, 392, 332
% (Analytically and numerically calculate the specific energy distribution in the disrupted star for different stellar types, which directly translates to the temporal behavior of the mass fallback rate)

\reference{} Lodato G., Rossi E. M., 2011, MNRAS, 410, 359

\reference{} Loeb, A., Ulmer, A. 1997, ApJ, 489, 573 

\reference{} Lynden-Bell D., Pringle J. E., 1974, MNRAS, 168, 603

\reference{} Mashhoon B., Hehl F. W., Theiss D. S., 1984, General Relativity and Gravitation, 16, 711

\reference{} Metzger B. D., Piro A. L., Quataert E., 2008, MNRAS, 390, 781

\reference{} Metzger B. D., Giannios D., Mimica P., 2012, MNRAS, 420, 3528

\reference{} Metzger B. D., Rafikov R. R., Bochkarev K. V., 2012, MNRAS, 423, 505

\reference{} Milosavljevi{\'c}, M., \& Phinney, E.~S.\ 2005, ApJ, 622, L93

\reference{} Montesinos Armijo M., de Freitas Pacheco J. A., 2011, ApJ, 736, 126

\reference{} Miller, J. M. \& G\"{u}ltekin, K., 2011, ApJ, 738, L13

\reference{} Narayan, R.; Igumenshchev, I. V.; Abramowicz, M. A., 2000, ApJ, 539, 798

\reference{} Narayan, R.; Piran, T.; Kumar, P., 2001, ApJ, 557, 949
% (CDAF can't produce GRBs because most matter escape to infinity instead of accreting to BH. NDAF can)

\reference{} Natarajan, P.; Pringle, J. E., 1998, ApJ, 506, L97
% (Hole-disk alignement for AGNs, for which $H/R \sim 10^{-3} - 10^{-2}$ and probably $< \alpha$. In this case, warps propagate is diffusion manner)

\reference{} Nelson, R. P.; Papaloizou, J. C. B., 2000, MNRAS, 315, 570
% (SPH simulation of hole-disk alignment for the $\alpha < H/R < 1$ regime where warps propagate in waves; calculate radius ($\sim 15 - 30 R_G$) out to which disk is aligned with hole's spin, considering pressure in the disk)

\reference{} Nelson, R. P.; Papaloizou, J. C. B., 1999, MNRAS, 309, 929
% (warps propagate in waves for $\alpha < H/R < 1$, and in diffusion for $H/R < \alpha < 1$)

\reference{} Papaloizou J. C. B., Pringle J. E., 1983, MNRAS, 202, 1181
% viscous warp propagation speed is v_r/(2 alpha^2).

\reference{} Pringle J. E., 1991, MNRAS, 248, 754
% (Time evolution of a disk around a NS; consider a angular momentum source term at the inner boundary of disk)

\reference{} Quataert E., Kasen D., 2012, MNRAS, 419, L1

\reference{} Ramirez-Ruiz E., Rosswog S., 2009, ApJ, 697, L77
% (Numerically studies dynamics of tidal disruption of a star by an IMBH in globular clusters; simulation shows mass returning rate is stable up to $t_{f}$.)

\reference{} Rees M. J., 1988, Nature, 333, 523

\reference{} Rees M. J., 1990, Science, 247, 817

\reference{} Reis R. C., Miller J. M., Reynolds M. T., G\"ultekin K., Maitra D., King A. L., Strohmayer T. E., 2012, Science, 337, 949

\reference{} Sakimoto P. J. \& Coroniti  F. V., 1981, ApJ 247, 19

\reference{} Saxton C. J., Soria R., Wu K., Kuin N. P. M., 2012, MNRAS, 422, 1625
% (Long-term X-ray variability analysis of Sw J1644 up to $t=2\times10^7$ s; quasi-period becomes larger at later epochs)

\reference{} Saxton, R. D.; Read, A. M.; Esquej, P.; Komossa, S.; Dougherty, S.; Rodriguez-Pascual, P.; Barrado, D., 2012, A\&A, 541, 106
% (The 3rd best TDE flare identified in X-ray band)

\reference{} Sorathia, K.~A., Krolik, J.~H., \& Hawley, J.~F.\ 2013, \apj, 768, 133

\reference{} Stone, J. M.; Gardiner, T. A.; Teuben, P.; Hawley, J. F.; Simon, J. B., 2008, ApJS, 178, 137

\reference{} Stone N., Loeb A., 2012, Phys. Rev. Lett., 108, 061302

\reference{} Stone, N.; Sari, R.; Loeb, A., 2012, arXiv:1210.3374

\reference{} Stone, J.~M. \& Norman, M.~L., 1992, ApJS 80, 753

\reference{} Strubbe L. E., Quataert E., 2009, MNRAS, 400, 2070

\reference{} Strubbe, L.~E., \& Quataert, E.\ 2011, \mnras, 415, 168

\reference{} Szuszkiewicz E., Miller J. C., 2001, MNRAS, 328, 36

\reference{} Taam, R. E.; Chen, X.; Swank, J. H., 1997, ApJ, 485, L83

\reference{} Tanaka, T., \& Menou, K.\ 2010, ApJ, 714, 404

\reference{} Tanaka T., 2011, MNRAS, 410, 1007

\reference{} Tchekhovskoy A., Metzger B. D., Giannios D., Kelley L. Z., 2013, arXiv:1301.1982

\reference{} Tout, C. A.; Pols, O. R.; Eggleton, P. P.; Han, Z. W., 1996, MNRAS, 281, 257

\reference{} Turner, N. J. \& Stone, J. M., 2001, ApJS 135, 95

\reference{} Ulmer A., 1999, ApJ, 514, 180

\reference{} van Velzen S., Farrar G. R., Gezari S., Morrell N., Zaritsky D., Ostman L., Smith M., Gelfand J., 2011, ApJ, 741, 73

\reference{} Wang F. Y., Cheng K. S.,  2012, MNRAS, 421, 908
% (Jet internal shock interpretation of the X-ray flares seen in SwJ-1644)

\reference{} Yuan, F.; Quataert, E.; Narayan, R., 2003, ApJ, 598, 301
% (Radiatively inefficient accretion flow in Sgr*. Assume $\dot{M}_acc \propto r^s$ following Blandfor & Begelman 1999)

\reference{} Zauderer B. A., et al., 2011, Nature, 476, 425

\reference{} Zauderer, B. A.; Berger, E.; Margutti, R.; Pooley, G. G.; Sari, R.; Soderberg, A. M.; Brunthaler, A.; Bietenholz, M. F., 2013, ApJ, 767, 152

\end{references}
\end{document}